\documentclass[manuscript]{aastex62}

\hypersetup{linkcolor=red,citecolor=cyan,filecolor=blue,urlcolor=magenta}
\usepackage{subfigure}
\usepackage{longtable}
\usepackage{rotating}
\usepackage{supertabular}

\begin{document}

\title{Active galactic nuclei with GeV activities and the PeV neutrino source candidate TXS 0506+056}

\correspondingauthor{Neng-Hui Liao, Hao-Ning He, Yi-Zhong Fan}
\email{liaonh@pmo.ac.cn, hnhe@pmo.ac.cn, yzfan@pmo.ac.cn}

\author[0000-0001-6614-3344]{Neng-Hui Liao}
\affiliation{Key Laboratory of Dark Matter and Space Astronomy, Purple Mountain Observatory, Chinese Academy of Sciences, Nanjing 210034, China}

\author{Yu-Liang Xin}
\affiliation{Key Laboratory of Dark Matter and Space Astronomy, Purple Mountain Observatory, Chinese Academy of Sciences, Nanjing 210034, China}

\author[0000-0002-6316-1616]{Yun-Feng Liang}
\affiliation{Key Laboratory of Dark Matter and Space Astronomy, Purple Mountain Observatory, Chinese Academy of Sciences, Nanjing 210034, China}

\author{Xiao-Lei Guo}
\affiliation{Key Laboratory of Dark Matter and Space Astronomy, Purple Mountain Observatory, Chinese Academy of Sciences, Nanjing 210034, China}
\affiliation{School of Astronomy and Space Science, University of Science and Technology of China, Hefei 230026, Anhui, China}

\author{Shang Li}
\affiliation{Key Laboratory of Dark Matter and Space Astronomy, Purple Mountain Observatory, Chinese Academy of Sciences, Nanjing 210034, China}

\author{Hao-Ning He}
\affiliation{Key Laboratory of Dark Matter and Space Astronomy, Purple Mountain Observatory, Chinese Academy of Sciences, Nanjing 210034, China}
\affiliation{Astrophysical Big Bang Laboratory, RIKEN, Wako, Saitama, Japan}

\author[0000-0001-5135-5942]{Qiang Yuan}
\affiliation{Key Laboratory of Dark Matter and Space Astronomy, Purple Mountain Observatory, Chinese Academy of Sciences, Nanjing 210034, China}

\author[0000-0002-8966-6911]{Yi-Zhong Fan}
\affiliation{Key Laboratory of Dark Matter and Space Astronomy, Purple Mountain Observatory, Chinese Academy of Sciences, Nanjing 210034, China}

\begin{abstract}
On 2017 September 22 the IceCube neutrino observatory detected a track-like, very-high-energy event (IceCube-170922A) that is spatially associated with TXS 0506+056, a quasar at a redshift of $z=0.3365$.  This source is characterized by the increased acitivies in a very wide energy range (from radio to TeV) during these days. To investigate the possible connection of the PeV neutrino emission with the GeV activity of blazars, in this work we select 116 bright sources and analyze their lightcurves and spectra. We focus on the sources displaying GeV activities. Among these blazars, TXS 0506+056 seems to be typical in many aspects but is distinguished by the very strong GeV activties. We suggest to search for neutrino outburst in the historical data of IceCube, as recently done for TXS 0506+056, from the directions of these more energetic and harder blazars with strong GeV activities.  
\end{abstract}

\keywords{galaxies: active -- gamma rays: galaxies -- blazars}

\section{Introduction} \label{sec:intro}
The discovery of astrophysical neutrino flux around PeV energies by IceCube \citep{2017JInst..12P3012A} is an important milestone in high energy astronomy \citep{2013Sci...342E...1I}. The neutrino is an ideal astronomical messenger since they travel undistorted from the sources and are therefore a valuable probe of the innermost regions of the energetic and enigmatic objects in the
cosmos. Since then, the accumulating observations of neutrino events from IceCube suggest that a significant fraction of the observed neutrinos are of extragalactic origin due to their isotropic distribution and reveal a flux of neutrinos with a total energy density comparable with that of the extragalactic $\gamma$ rays observed by {\it Fermi}-LAT \citep{2014PhRvL.113j1101A,2015PhRvL.115h1102A,2016PhRvL.116o1105A}.

Blazars are an extreme subtype of Active Galactic Nuclei (AGNs), dominating the extragalactic $\gamma$-ray sky \citep{2015ApJS..218...23A,2016ARA&A..54..725M}. Emissions from their strong collimated jets are overwhelming due to relativistic beaming effects and hence blazars are characterized by the luminous and highly variable broadband continuum emissions \citep[e.g.,][]{BR78,1997ARA&A..35..445U}. Blazars are traditionally divided into flat spectrum radio quasars (FSRQs) and BL Lacertae objects (BL Lacs) based on their optical spectra. Spectral energy distribution of these jetted AGNs generally exhibits a two-bump structure in log$\nu$F$_\nu-$log$\nu$ representation and the low energy bump is widely believed to be contributed by synchrotron emission. Therefore, blazars are also classified as low-synchrotron-peaked sources (LSPs, $\nu^{syn}_{peak} <$ $10^{14}$ Hz), intermediate-synchrotron-peaked sources (ISPs, $10^{14}$ Hz $< \nu^{syn}_{peak} <$ $10^{15}$ Hz), and high-synchrotron-peaked sources ($\nu^{syn}_{peak} >$ $10^{15}$ Hz)\citep{2015ApJ...810...14A}. However, the origin of the other bump extending to $\gamma$-ray regime is still under debate. On one hand, the leptonic radiation model \citep[e.g.,][]{1992ApJ...397L...5M,1993ApJ...416..458D,1994ApJ...421..153S,2000ApJ...545..107B} in which the $\gamma$ rays are from inverse Compton scattering of soft photons by the same population of electrons that emit the synchrotron emission, can naturally describe the tightly connected multiwavelength variability of blazars \citep[e.g.,][]{2010Natur.463..919A,2014ApJ...783...83L}. On the other hand, observational phenomena, like the ``orphan'' $\gamma$-ray flare \citep[e.g.,][]{2005ApJ...621..176B}, support the hadronic scenario that $\gamma$ rays and neutrinos are produced via  interactions of high-energy protons with gas (i.e., the pp-interactions) in the jets \citep{2002A&A...382..829S} or in interactions of protons with internal \citep{1995APh.....3..295M} or external \citep{2001PhRvL..87v1102A} photon fields (p$\gamma$-interactions). 

Among the various possibilities of potential extragalactic sources of neutrinos (see \cite{2015RPPh...78l6901A} for a review), including star-forming galaxies \citep[e.g.,][]{2006JCAP...05..003L}, gamma-ray bursts \citep[e.g.,][]{1997PhRvL..78.2292W}, galaxy clusters~\citep[e.g.,][]{1997ApJ...487..529B} and so on, blazars are believed to be promising sources\citep[e.g.,][]{2015MNRAS.451.1502T,2016ApJ...831...12H}. 
This kind of sources also contribute to the majority of extragalactic $\gamma$-ray background which is consistent with the measured neutrino flux level \citep{2013PhRvD..88l1301M}. 
Meanwhile, a correlation between cosmic neutrinos and blazar catalogs has been argued \citep{2016MNRAS.457.3582P}. Moreover, since strong $\gamma$-ray flares of blazars have been frequently detected, a spatial association combined with a coincidence in time with a flaring blazar may represent a smoking gun for the origin of the IceCube flux. A coincidence between a 2 PeV neutrino event and the blazar PKS B1424$-$418 provides an interesting hint in this context \citep{2016NatPh..12..807K}. 

On 2017 Spetmber 22, the IceCube neutrino observatory (hereafter IceCube) detected a track-like, very-high-energy event with a high probability of being of astrophysical origin \citep{GCN21916}. Inside the error region of the neutrino event, there is a {\it Fermi}-LAT source, TXS 0506+056 \citep{Ajello2017}. Interestingly, the GeV emission is found to be at high state \citep{ATel10791}. Enhanced emission are also found in radio, optical, X-ray and TeV bands \citep{ATel10791,ATel10845,ATel10794,ATel10817}. Therefore, TXS 0506+056 is a promising PeV neutrino source candidate \citep{IceCube2018b,IceCube2018a}. Motivated by the coincidence of the GeV activity with the PeV neutrino emission, in this work we analyze the {\it Fermi}-LAT data \citep{Fermi2009,Fermi2013} of some bright AGNs and identify the GeV flares.
We then compare the properties of these bright AGNs displaying strong GeV activities with TXS 0506+056 to look for possible indication of the high energy neutrino sources. This work is organized as follows: in Section 2 we introduce our sample and the data analysis. The results are reported in Section 3. We summarize our results with some discussions in Section 4.

\section{THE SAMPLE AND DATA ANALYSES} \label{sec:data}
\subsection{The Sample}
Blazars included in the {\it Fermi}-LAT monitored source list\footnote{https://fermi.gsfc.nasa.gov/ssc/data/access/lat/msl\_lc/} are chosen for this study, in which the sources are selected if their instantaneous weekly fluxes are above $\rm 10^{-6}$ ph $\rm cm^{-2}$ $\rm s^{-1}$. These sources represent the brightest and the most variable ones among the {\it Fermi}-LAT blazars. Another advantage is that the sources are randomly distributed in the sky since {\it Fermi}-LAT performs an all-sky survey. Several sources (e.g. Fermi J2007$-$2518) are excluded in our analysis because there are only a few flux data points in the preliminary weekly light curves provided by the {\it Fermi} collaboration. Meanwhile, two sources (i.e. NRAO~190 and B2~1144+40) are not considered due to the ambiguous association relationship between the $\gamma$-ray source and its low-frequency counterpart. Therefore, our sample consists of total 116 blazars, including 81 FSRQs, 22 BL Lacs and 13 sources with unknown optical spectral property (BZU), according to BZCAT \footnote{http://www.asdc.asi.it/bzcat/} \citep{2009A&A...495..691M}. Or alternatively, there are 96 LSPs, 10 ISPs and 8 HSPs, according to the third {\it Fermi}-LAT AGN catalog (3LAC, \citealt{2015ApJ...810...14A}), together with 2 sources lacking of relative information. The redshift distribution of our sample is between 0.031 (Mrk 421) and 2.852 (PKS 0438$-$43), and there are seven sources (either BL Lacs or BZUs) lacking redshift imformation. The basic information of the sources in our sample are listed in Table \ref{Table 1}.

\subsection{{\it Fermi}-LAT Data reduction} \label{subsec:data-reduction}
In the analysis, the latest Pass 8 version of the {\it Fermi}-LAT data \citep{Fermi2013} with ``{\tt Source}'' event class ({\tt evclass} = 128 \& {\tt evtype} = 3) are selected, recorded from May 1, 2010 (Mission Elapsed Time 294364802; the time of the operation for the full IceCube detector)  to May 1, 2018 (Mission Elapsed Time 546825605)\footnote{Note there is $\sim$ 20 day data gap for {\it Fermi}-LAT around March 2018.}. We choose the events within a $10^\circ$ region of interest (ROI) with energies between 100 MeV and 500 GeV in this analysis. In order to reduce the contamination from the Earth Limb, the events with zenith angles larger than $90^\circ$ are excluded. In addition, the entire data set is filtered with {\tt gtmktime} to obtain high-quality data in the good time intervals, with the expression recommended by the LAT team, namely {\tt (DATA\_QUAL$>$0)\&\&(LAT\_CONFIG==1)}.
The data are analyzed with the standard LAT analysis software, {\it ScienceTools} version {\tt v10r0p5}\footnote {http://fermi.gsfc.nasa.gov/ssc/data/analysis/software/}, available from the Fermi Science Support Center, and the ``P8R2{\_}SOURCE{\_}V6'' instrumental response functions are adopted.  

First, we perform the {\tt unbinned} likelihood analysis with {\tt gtlike} to extract the global flux and spectral parameters of the target source. For the background subtraction, the Galactic diffuse emission and the isotropic diffuse emission are modeled by {\tt gll\_iem\_v06.fits} and {\tt iso\_P8R2\_SOURCE\_V6\_v06.txt}, which can be found from the Fermi Science Support Center \footnote{http://fermi.gsfc.nasa.gov/ssc/data/access/lat/BackgroundModels.html}. Meanwhile, all sources in the preliminary LAT 8-year Point Source List (FL8Y\footnote{https://fermi.gsfc.nasa.gov/ssc/data/access/lat/fl8y/}) within a radius of $15^\circ$ from the ROI center are included in the source model. For convenient comparison between different sources, the spectral template of each target source is set to a power-law function (i.e. $dN/dE \propto E^{-\Gamma}$, where $\Gamma$ is the spectral photon index). And during the fitting analysis, the normalizations and spectral parameters of all sources within a distance of $10^\circ$ from the ROI center, together with the normalizations of the two diffuse backgrounds, are left free. The significance of the target source can be quantified by the test statistic (TS) value, which is defined as TS = $-$2$~$ln($L_0/L$)\citep{1996ApJ...461..396M}, where $L_0$ and $L$ are the maximum likelihood values of the null hypothesis and the tested model including the target source. The fittings are demanded to converge (i.e. ``fit quality = 3'') to make sure the results are valid. The best-fit results for all target sources in the sample are summarized in Table \ref{Table 1}.

In the temporal analysis, we divide the data into 417 equal time bins and repeat the likelihood fitting for each time bin to extract a weekly light curve for each target. Considering the targets studied here are the brightest $\gamma$-ray sources in the extragalactic sky,  we fix the spectral indexes of all background sources to the global fitting values, and only free the normalization parameters of background sources within $10^\circ$ from the targets and two diffuse backgrounds. During the fitting analysis for each time bin, the weak background sources (i.e., TS $<$ 5) are removed from the source model. Note that for any time bin in which the TS value of target source is smaller than 9, the 95\% upper limit is calculated instead. 

\section{Results} \label{sec:resul}
\subsection{Global properties}
The analysis results of the entire 8-year LAT data are summarized in Table \ref{Table 1} as well. We also plot the photon index versus the photon flux and the $\gamma$-ray apparent luminosity (from 0.1 to 500~GeV, with handled $k$-correction) diagrams, see Fig. \ref{fig:lg}. The average photon indexes of FSRQs, BL Lacs and BZUs in our sample are 2.40$\pm$0.17, 2.04$\pm$0.18 and 2.29$\pm$0.12, respectively, which are in agreement with the results in 3LAC \citep{2015ApJ...810...14A}. Note that there is only one FSRQ, VER 0521+211, whose photon index is smaller than 2. As shown in Fig. \ref{fig:lg} and Fig. \ref{fig:f-l}, FSRQs are generally brighter and more luminous than BL Lacs, and TXS 0506+056 appears to be typical among other BL Lacs. The spectra of FSRQs are softer than that of BL Lacs, which may lower the neutrino detection possibility. 

\subsection{Temporal Behaviors}
\subsubsection{TXS 0506+056}
\begin{figure}
\centering
\includegraphics[width=0.45\columnwidth]{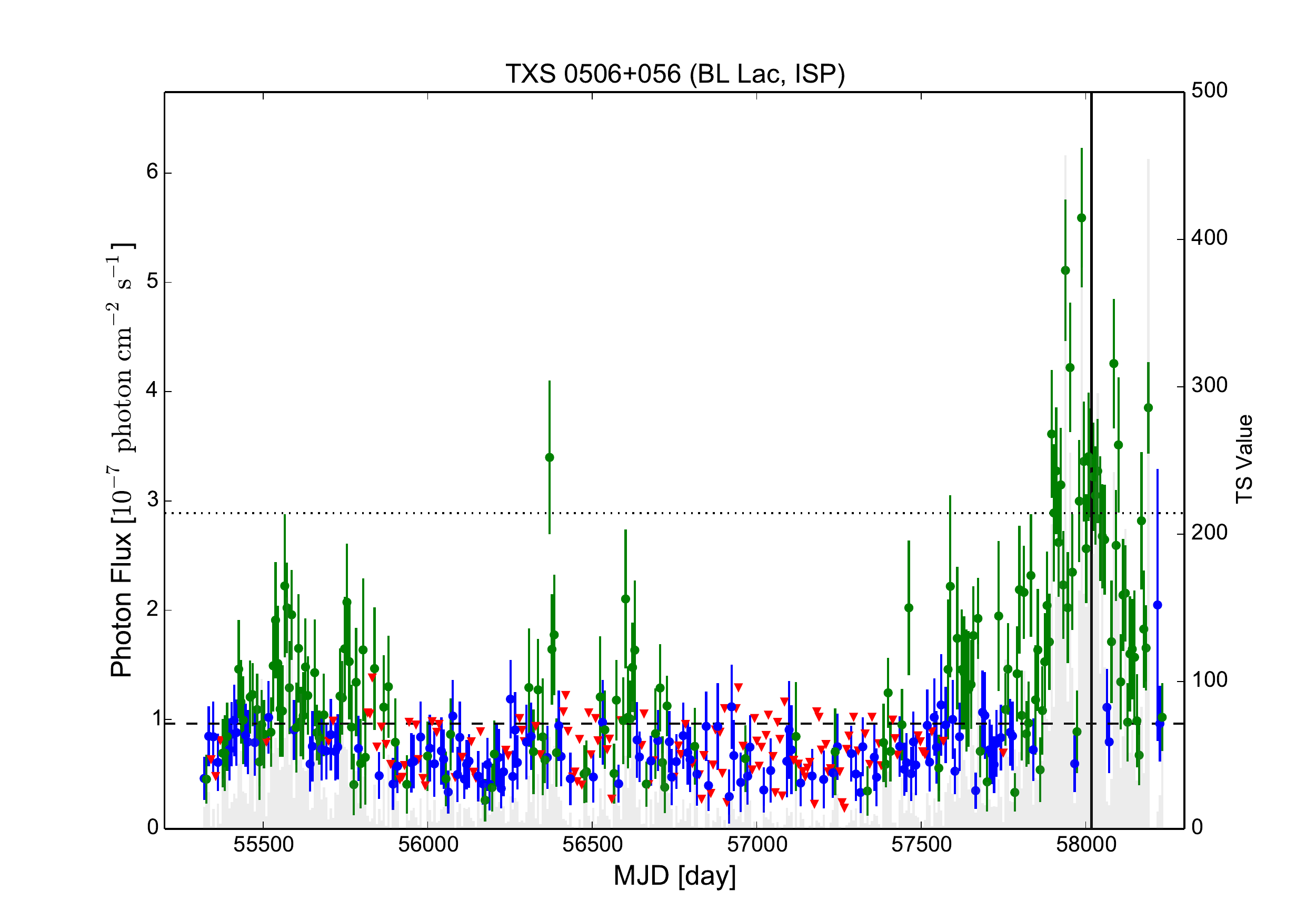}
\includegraphics[width=0.45\columnwidth]{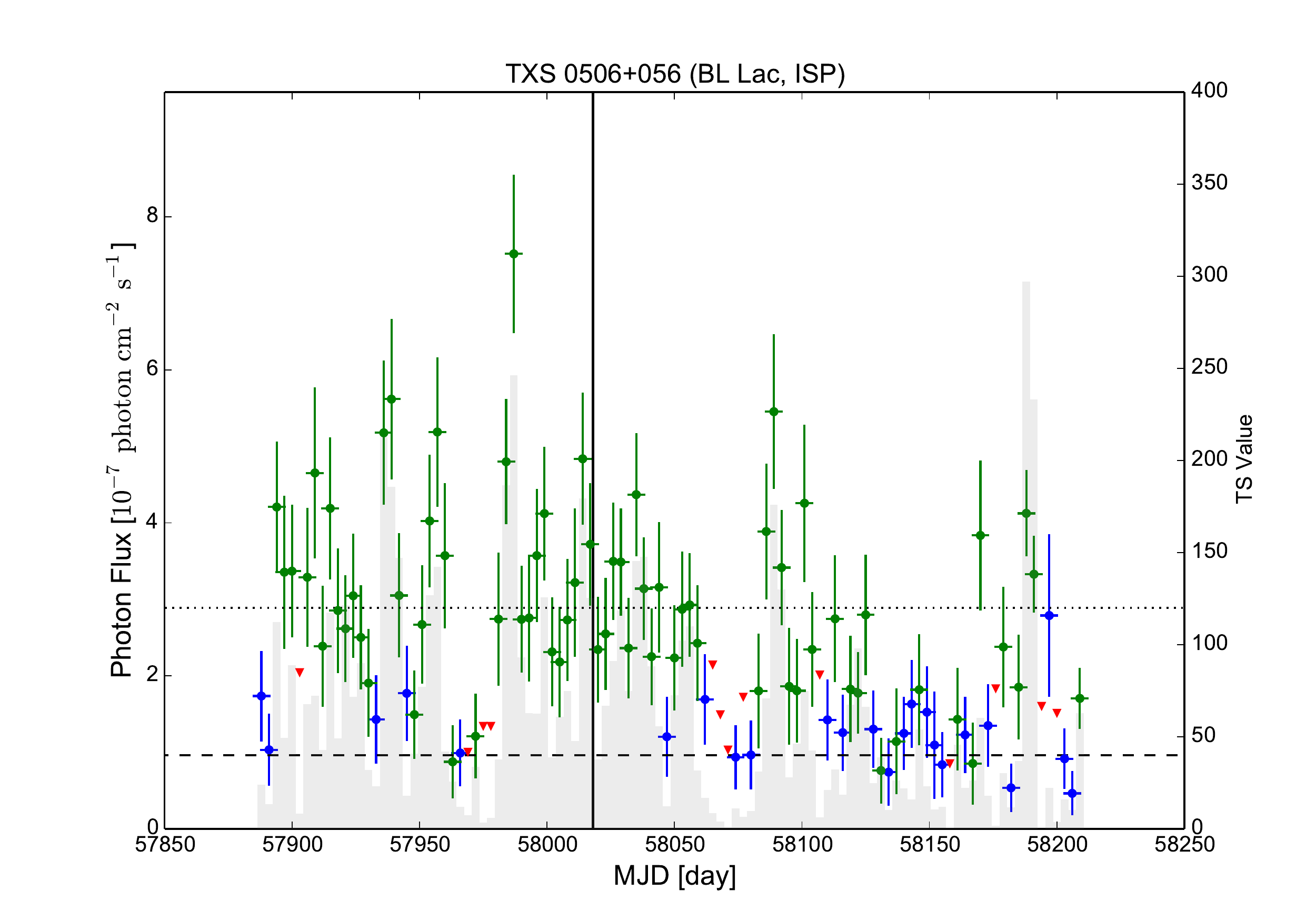}
\caption{The weekly $\gamma$-ray light curve from MJD 55317 to MJD 58239 (left) and the 3-day time bin $\gamma$-ray light curve from MJD 57888 to MJD 58239 (right) of TXS 0506+056.
For any time bin in which the TS value of TXS 0506+056 is larger than 25, the photon flux is derived by free the spectral index of TXS 0506+056, which is shown as the green dot.
The blue dots are for the time bin with $9<{\rm TS}<25$, and the fluxes in these bin are derived by keeping spectral indexes of all sources including TXS 0506+056 fixed. The red triangles are the 95\% upper limits for the time bins with TS values of TXS 0506+056 smaller than 9. The gray dashed line is
the average flux of TXS 0506+056 and the three-time average flux is shown as the black dotted line. The black solid line denotes the arrival time of the neutrino event detected by IceCube.}
\label{fig:0506wlc}
\end{figure}

The weekly $\gamma$-ray light curve of TXS 0506+056 is presented in the left panel of Fig. \ref{fig:0506wlc}. Before MJD 57855, the source maintains at a relatively quiescent state, with only a few bins whose fluxes reach roughly three times of the 8-year averaged flux, 9.8$\times 10^{-8}$ ph $\rm cm^{-2}$ $\rm s^{-1}$. However, since then, a strong $\gamma$-ray flare has appeared, with a peak flux of $(5.8\pm0.6)$ $\times 10^{-7}$ ph $\rm cm^{-2}$ $\rm s^{-1}$ which is nearly 6 times of the 8-year averaged flux. The corresponding peak apparent $\gamma$-ray luminosity of the bin is $\sim$ $1.5\times10^{47}$ erg $\rm s^{-1}$, adopting a redshift of 0.3365 \citep{2018ApJ...854L..32P}. Meanwhile, the photon index then is 2.05$\pm$0.07, suggesting no significant spectral variability than the average value, 2.07$\pm$0.01. The  
flare peaked on MJD 57981 (16th Aug. 2017), about a month before the arrival time of the IceCube neutrino event. A further 3-day $\gamma$-ray light curve (the right panel of Fig. \ref{fig:0506wlc}) shows that several sub-flares constitute this activity phase. Maybe there is a weak flare coincident with the neutrino event, but the error bars are relatively large. 

\subsubsection{Comparison between TXS 0506+056 and {\it Fermi}-LAT bright blazars}
\begin{figure}
\centering
\includegraphics[width=0.45\columnwidth]{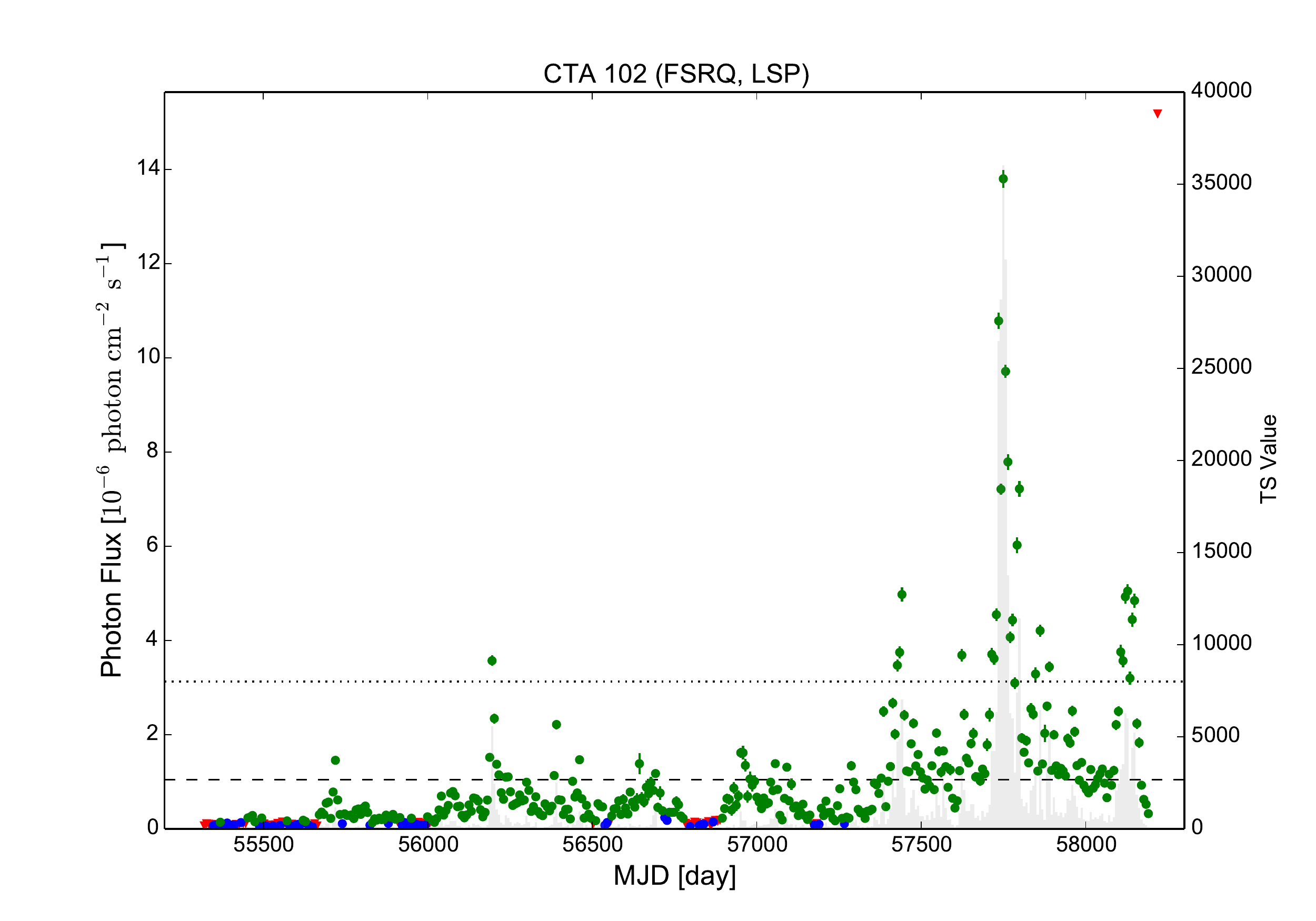}
\includegraphics[width=0.45\columnwidth]{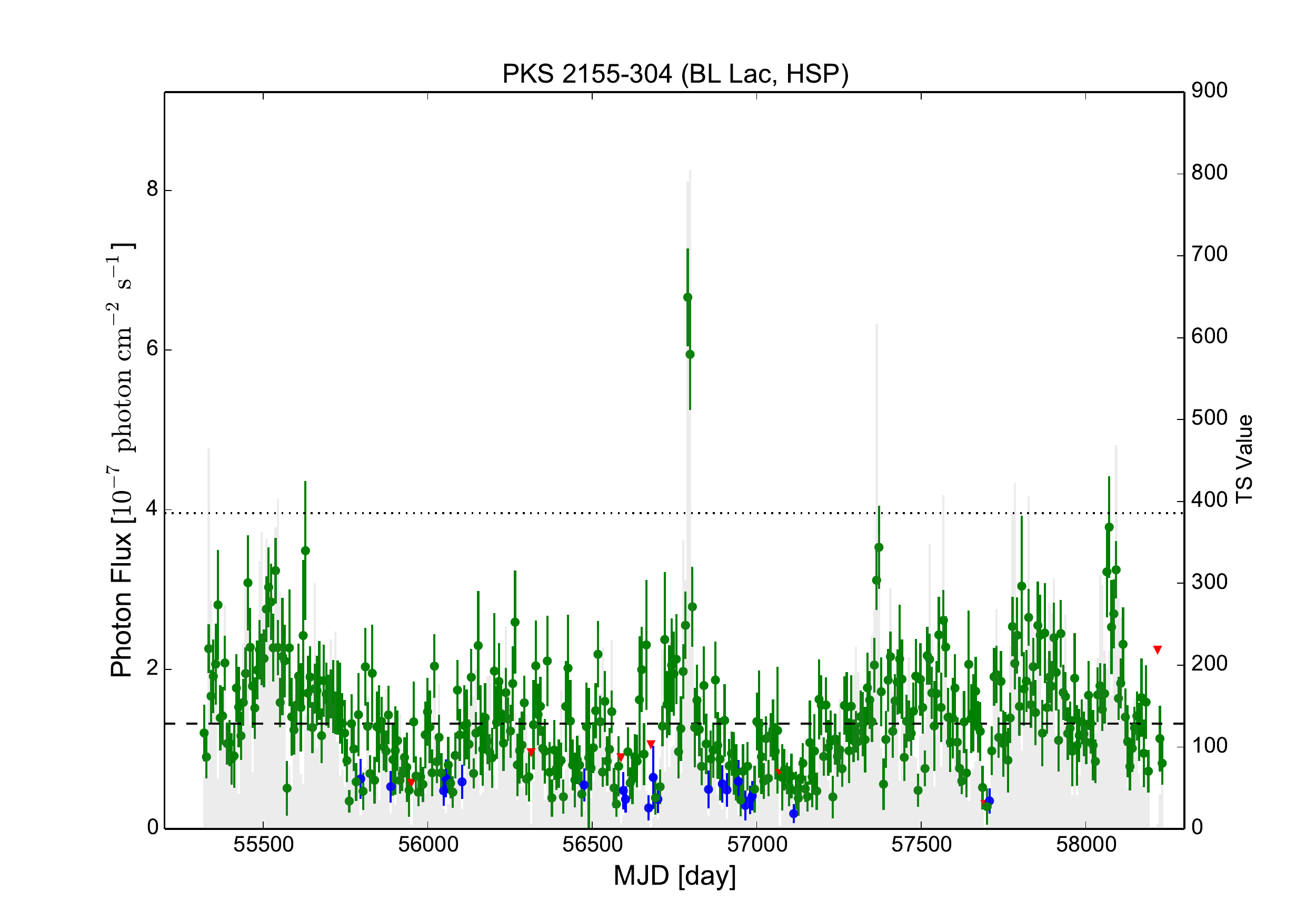}
\includegraphics[width=0.45\columnwidth]{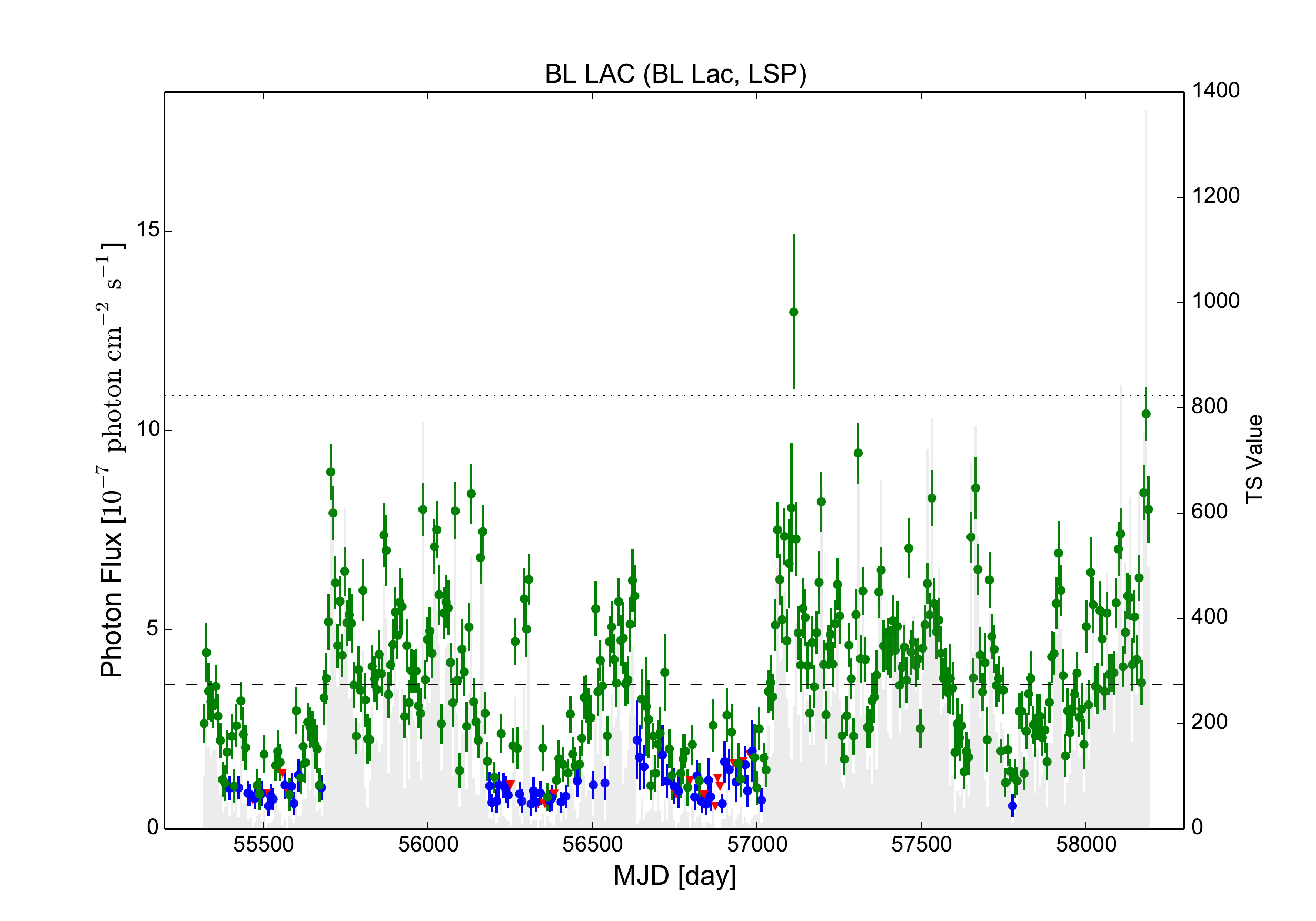}
\includegraphics[width=0.45\columnwidth]{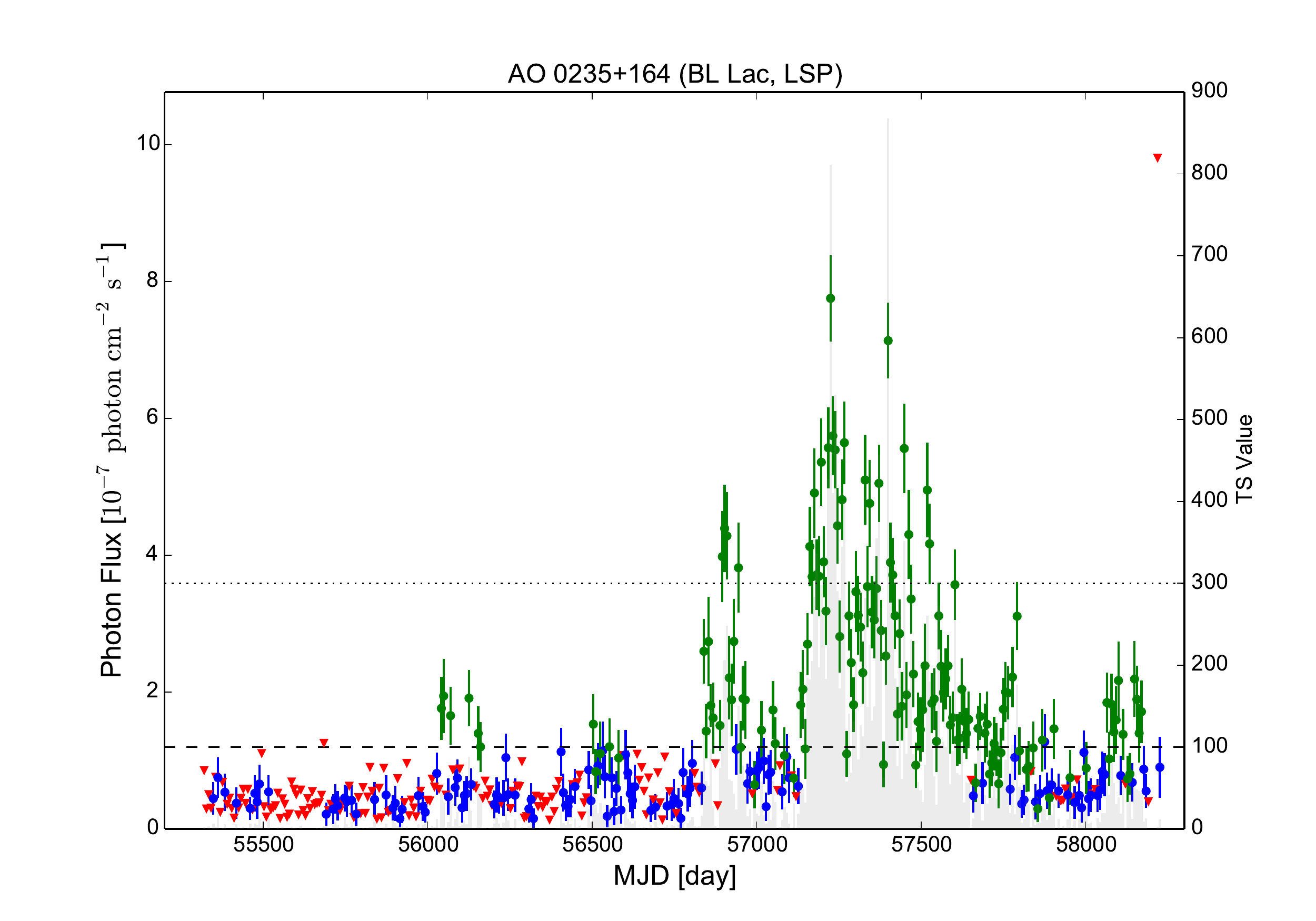}
\includegraphics[width=0.45\columnwidth]{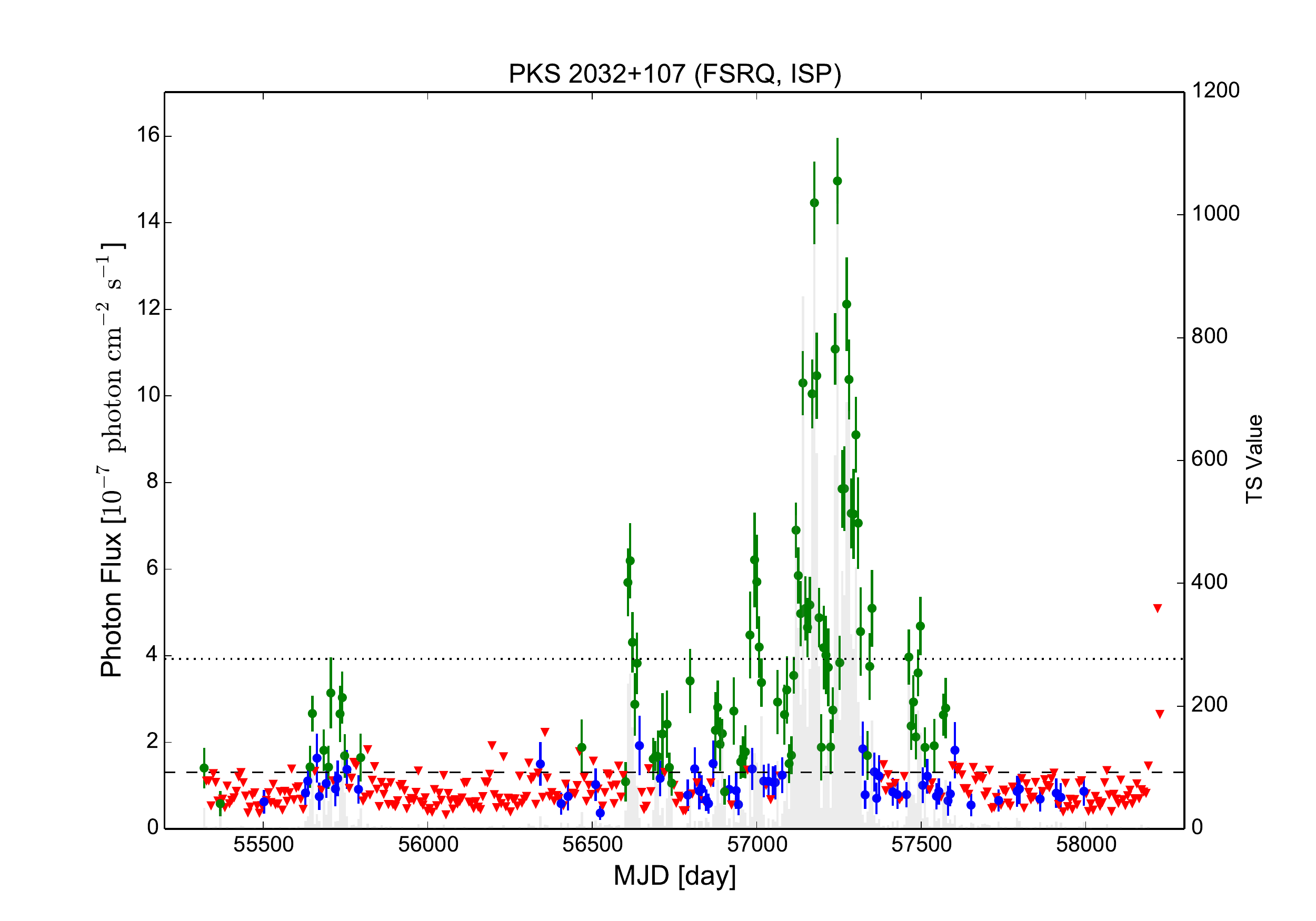}
\includegraphics[width=0.45\columnwidth]{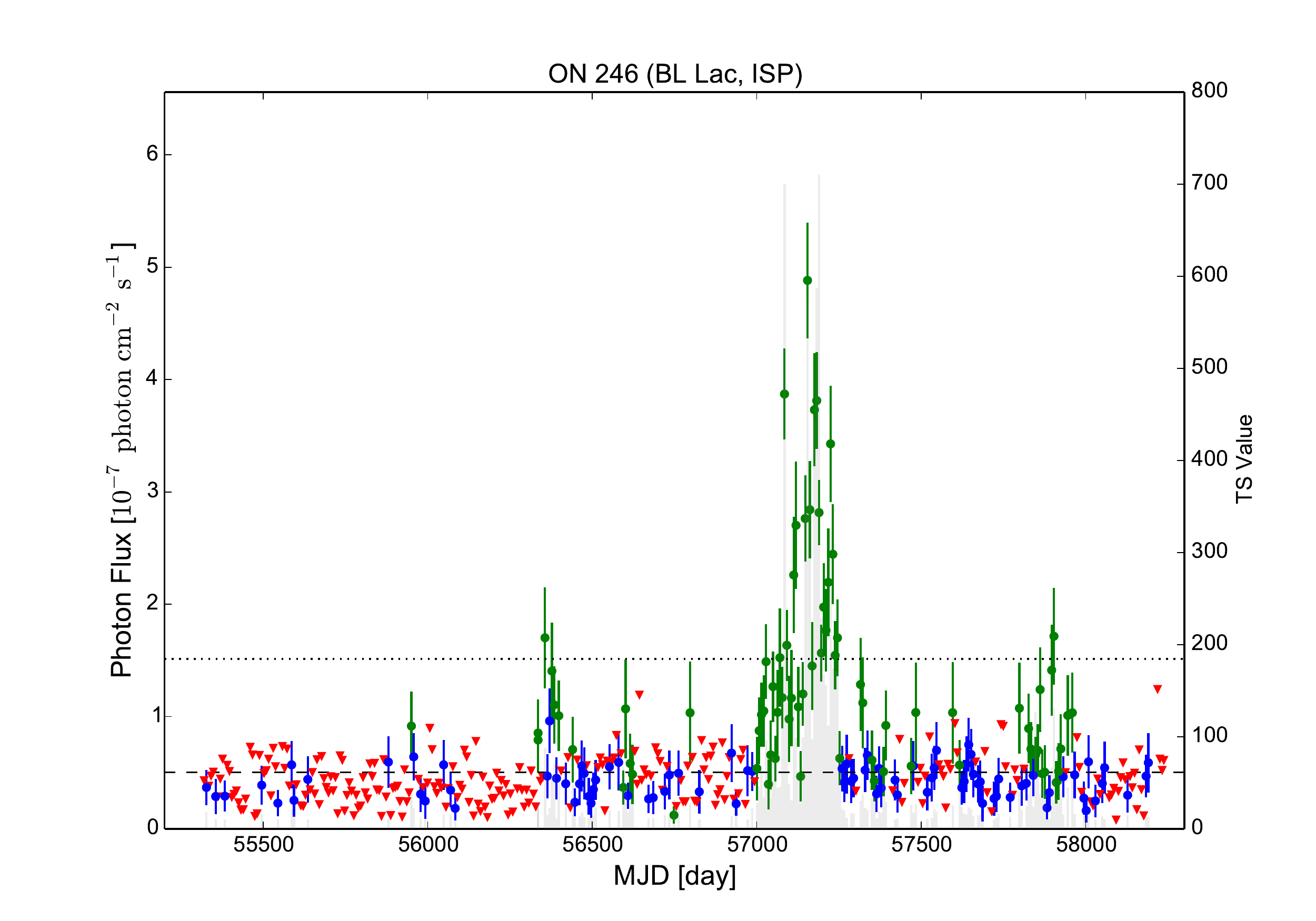}
\caption{Same as Fig. \ref{fig:0506wlc} but for different target sources. The name of the source and its type are marked as the title of each panel.}
\label{fig:others-wlc}
\end{figure}

In 3LAC, about 69\% FSRQs are found to be significantly variable, however, this fraction is down to 23\% for BL Lacs \citep{2015ApJ...810...14A}. Together with the fact that the former kind is generally brighter than the later one, it is not surprised that the vast majority of our sources are FSRQs. One famous case is CTA 102 \citep{cta102fermi,cta102dampe}, whose daily LAT $>$~100~MeV flux is up to $\rm 10^{-5}$ $\rm cm^{-2}$ $\rm s^{-1}$, also see Fig. \ref{fig:others-wlc} for its weekly light curve. In addition to the brightness, its large variability amplitude is also remarkable. The weekly peak flux is more than one order of magnitude higher than the averaged flux. Examples of light curves of other subtypes are also shown in Fig. \ref{fig:others-wlc}. Although most of our FSRQs are LSPs, there are several ISP FSRQs. No significant difference is found between their weekly $\gamma$-ray light curves. It is interesting to see that the variability amplitude for FSRQs is generally higher than HSP BL Lacs. However, for ISP BL Lacs and LSP BL Lacs, their $\gamma$-ray variability could be as violent as FSRQs. Worthy to note that there is one source, ON 246, which is also a ISP BL lac similar with TXS 0506+056. There is also a strong $\gamma$-ray flare with peak flux of 5 $\times 10^{-7}$ ph $\rm cm^{-2}$ $\rm s^{-1}$ for ON 246.

Since the neutrino event IceCube-170922A arrived when TXS 0506+056 was flaring in GeV band, we adopt $f_{\gamma}^{accu}/f_{\gamma}^{aver}$ as an indicator to qualify the comparison between TXS 0506+056 and other {\it Fermi} bright blazars. $f_{\gamma}^{accu}$ is defined as an accumulated photon flux of the {\it flares}, which should have fluxes brighter than the average ($f_{\gamma}^{aver}$) by a factor of 3. The division of $f_{\gamma}^{aver}$ is to eliminate the influence caused by a wide distribution of flux level for different sources. Therefore, $f_{\gamma}^{accu}/f_{\gamma}^{aver}$ reflects the intensity of GeV activity of a source (a large $f_{\gamma}^{accu}/f_{\gamma}^{aver}$ arises if a few flares have fluxes much higher than $f_{\gamma}^{aver}$ or alternatively there are intense flare activities). If the neutrino is indeed tightly connected to the flare event of blazars, sources with high $f_{\gamma}^{accu}/f_{\gamma}^{aver}$ value and hard $\gamma$-ray spectra are preferred to be identified as neutrino sources. In the $f_{\gamma}^{accu}/f_{\gamma}^{aver}-\Gamma_{\gamma}^{aver}$ plot, as shown in Fig. \ref{fig:foc}, TXS 0506+056 is distinguished by the high $f_{\gamma}^{accu}/f_{\gamma}^{aver}$ (i.e., the very strong GeV activities). Together with the declination of TXS 0506+056 (as seen in Fig. \ref{fig:Dec}; please note that the IceCube observatory has better sensitivity for sources with declination close to 0 \citep{IceCube2014}), it may help to explain why TXS 0506+056 is the first blazar associated
with a significant neutrino excess.  
We have also investigated the distribution of the accumulated isotropic-equivalent energy of the flare emission (i.e., $E_{\gamma}^{accu}$, with proper $k-$correction). In Fig.\ref{fig:L-hs} we show the diagram of $E_{\gamma}^{accu}-\Gamma_{\gamma}^{accu}$, where $\Gamma_{\gamma}^{accu}$ is the powerlaw index of the spectrum of the accumulated flare emission. In such a plot, TXS 0506+056 is not distinct. Finally we study the change of the hardness of the blazar emission in the flare phase. We define a new parameter $\Delta\Gamma_{\gamma}=\Gamma_{\gamma}^{aver}-\Gamma_{\gamma}^{accu}$. In the plot of $E_{\gamma}^{accu}-\Delta\Gamma_{\gamma}$ (see Fig. \ref{fig:Delta_gamma}), TXS 0506+056 is similar with other blazars, too. Therefore, TXS 0506+056 seems to be a normal blazar in many ways except for its very strong GeV activities. 
Dedicated neurtrino searches in the directions of some bright blazars with strong GeV activities are encouraged,
to check whether these sources are also important neutrino sources.

\section{Summary} \label{sec:diss}

\begin{figure}
\centering
\includegraphics[width=0.45\columnwidth]{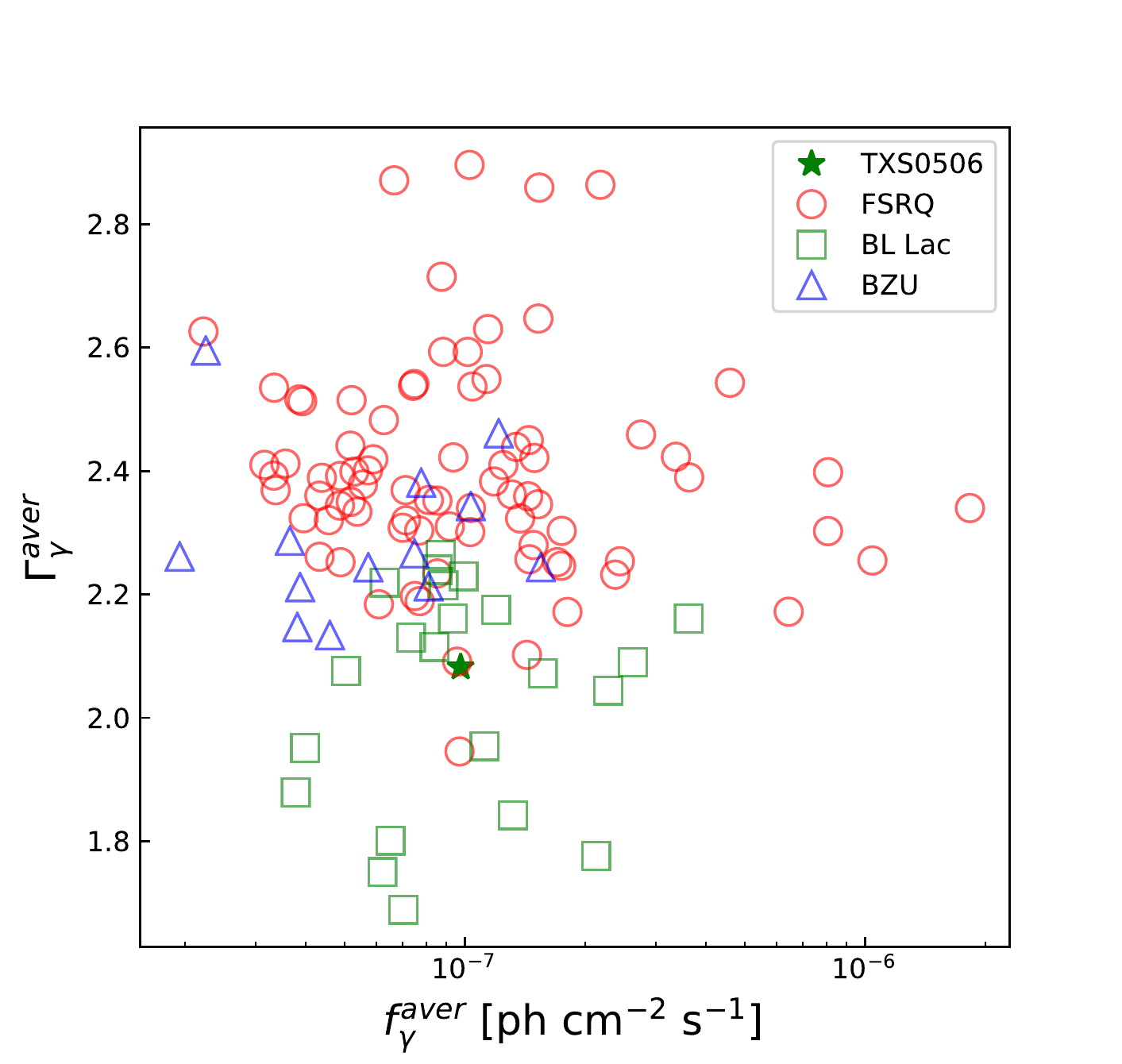}
\includegraphics[width=0.45\columnwidth]{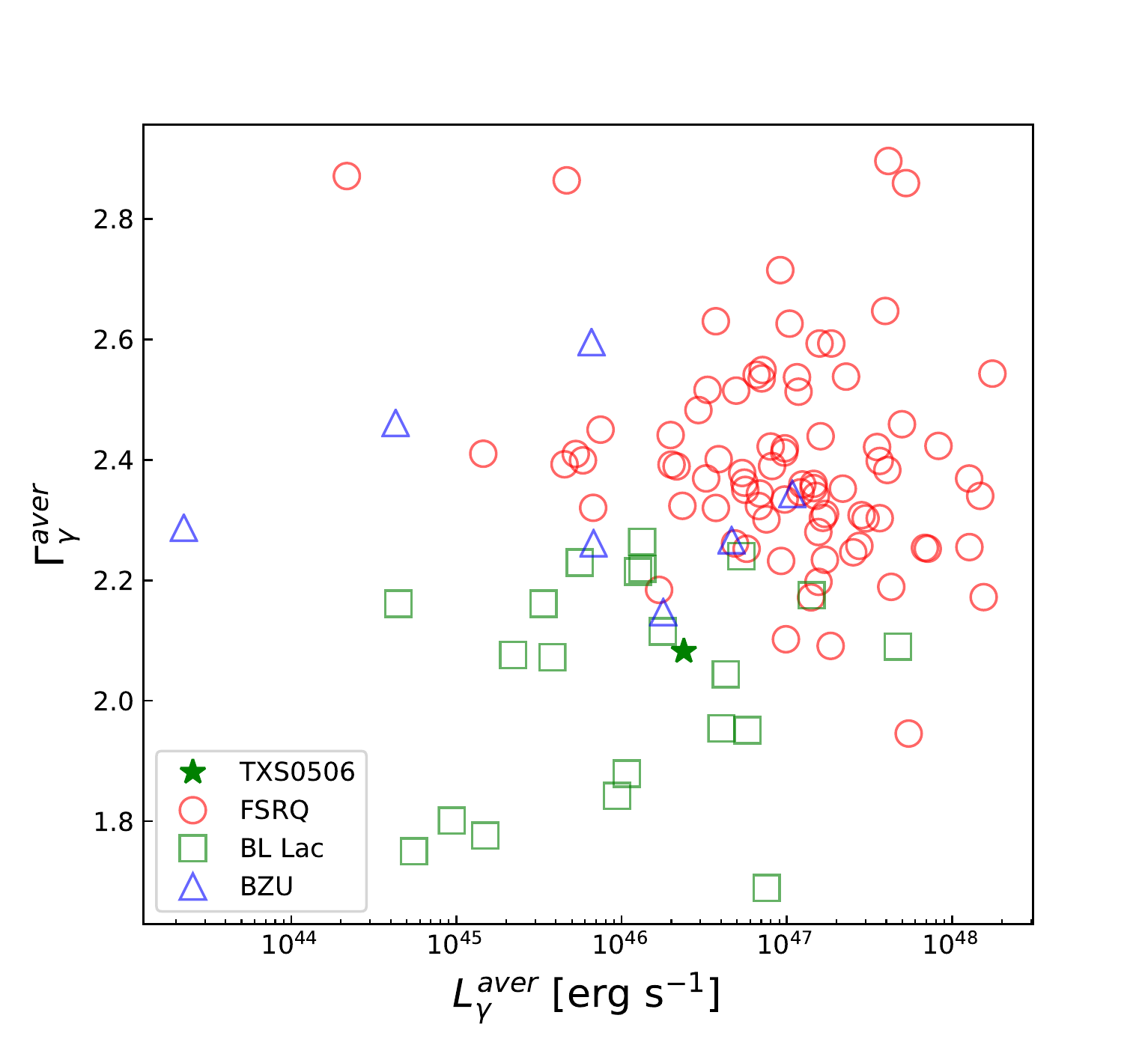}
\caption{TXS 0506+056 in the $L_{\gamma}^{aver} - f_{\gamma}^{aver}$ (left) and $L_{\gamma}^{aver} - \Gamma_{\gamma}^{aver}$ (right) diagrams. The superscript $aver$ means that they are average quantities for the whole 8-year data. The red circles, green squares and blue triangles represent the FSRQs, BL Lacs and the blazars of unknown type (BZU), respectively. TXS 0506+056 is marked as a green pentagram.}
\label{fig:lg}
\end{figure}

\begin{figure}
\centering
\includegraphics[width=0.8\columnwidth]{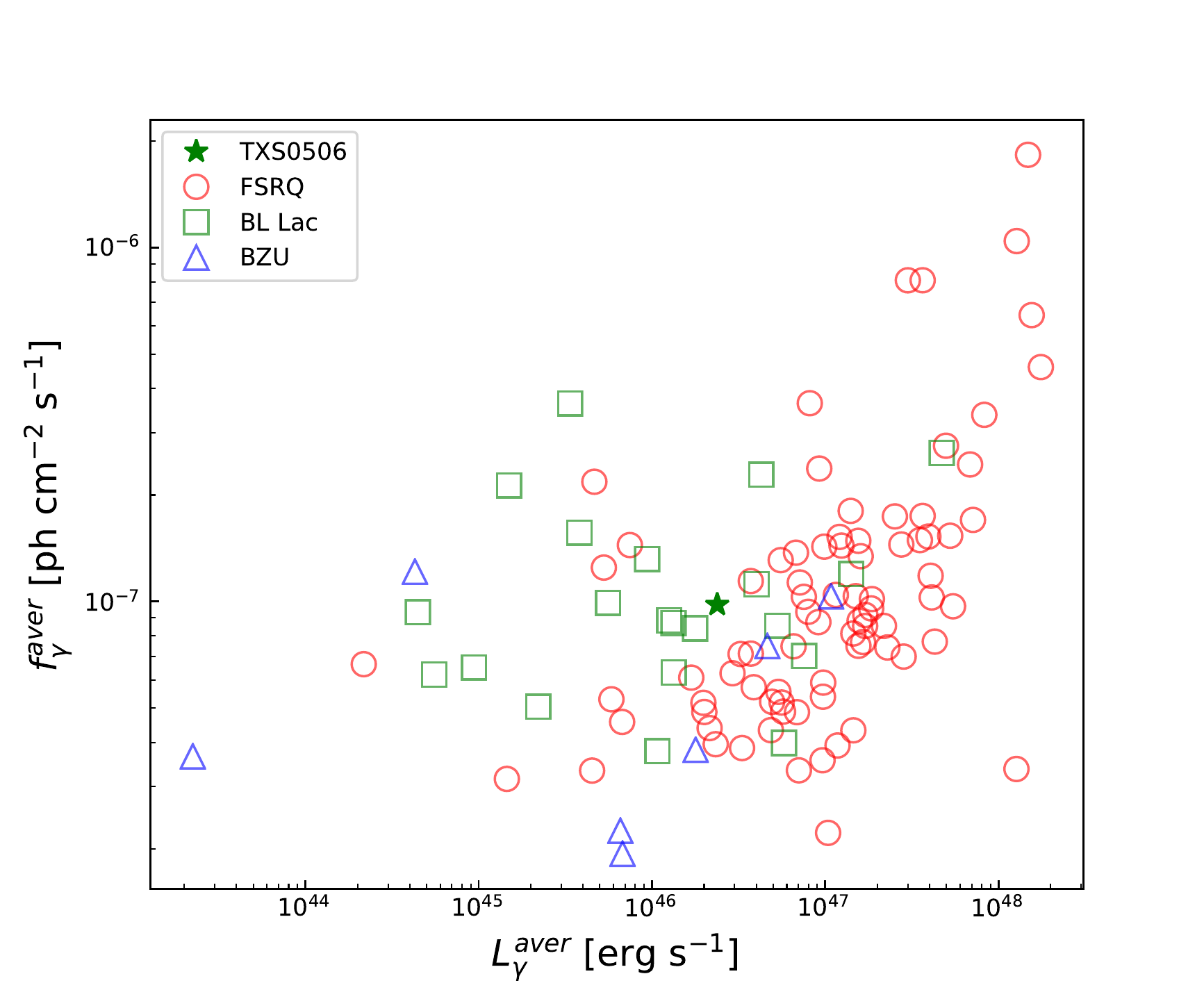}
\caption{TXS 0506+056, marked as the green pentagram, in the $L_{\gamma}^{aver}-f_{\gamma}^{aver}$ diagram. The legends are as the same of Fig \ref{fig:lg}.}
\label{fig:f-l}
\end{figure}

\begin{figure}
\centering
\includegraphics[width=0.8\columnwidth]{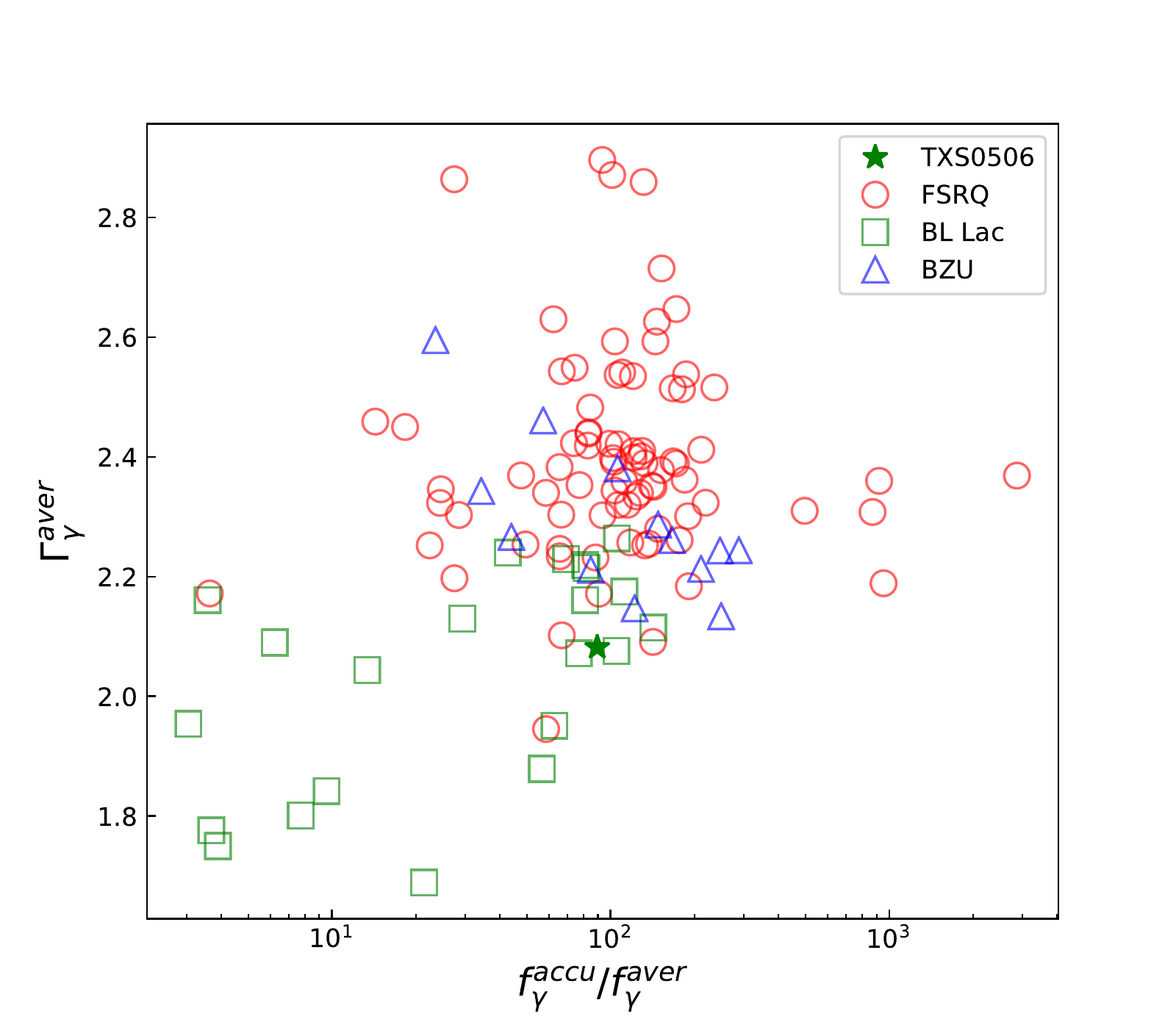}
\caption{The $f_{\gamma}^{accu}/f_{\gamma}^{aver}-\Gamma_{\gamma}^{aver}$ diagram, where $f_{\gamma}^{accu}$ represents the photon flux accumulated over the time bins with flux at least three times larger than $f_{\gamma}^{aver}$. TXS 0506+056 is marked as the green pentagram (same legend as for Fig. \ref{fig:lg}).}
\label{fig:foc}
\end{figure}

\begin{figure}
\centering
\includegraphics[width=0.8\columnwidth]{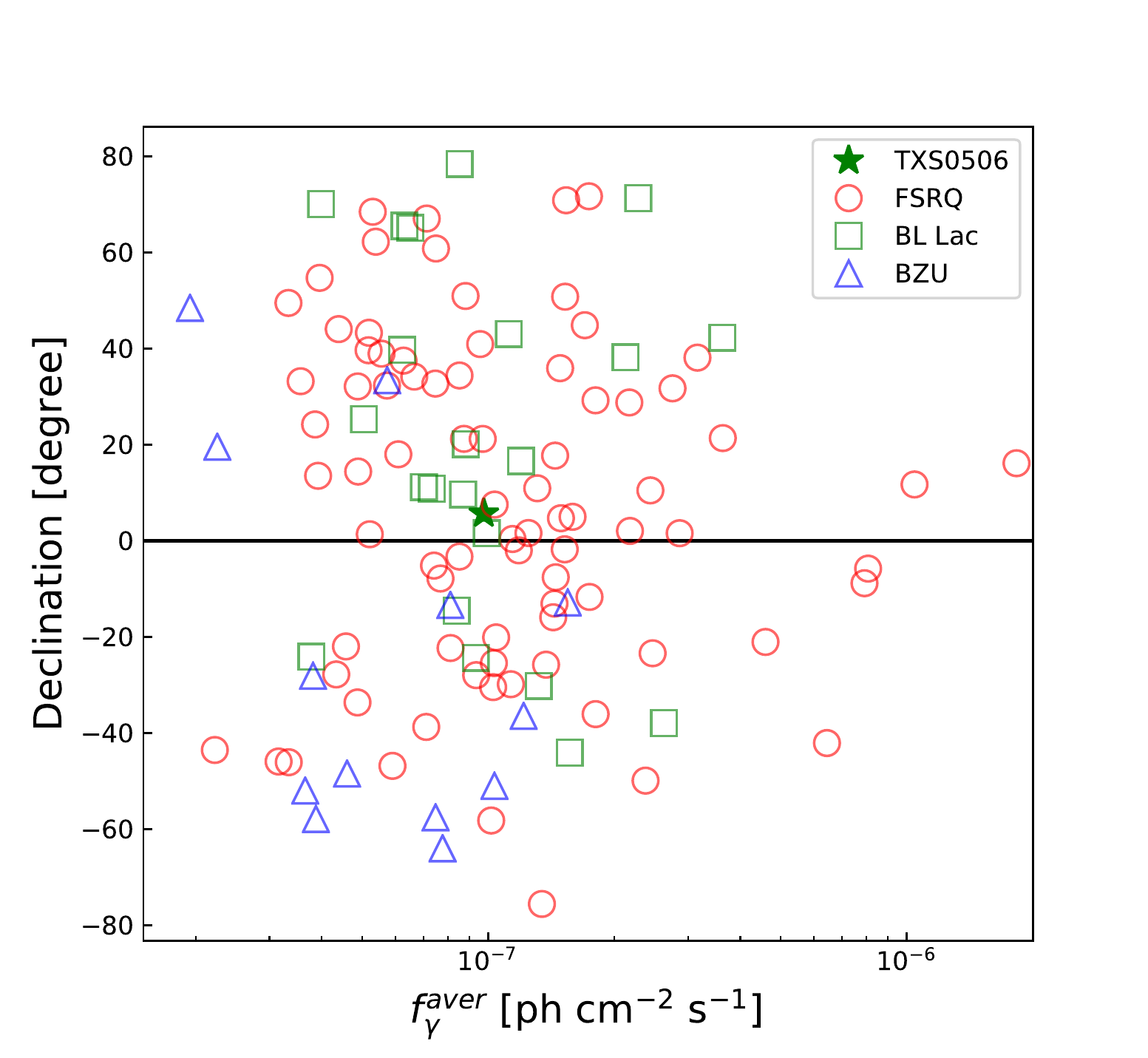}
\caption{The $Declination - f_{\gamma}^{aver}$ diagram. TXS 0506+056 is marked as the green pentagram (same legend as for Fig. \ref{fig:lg}).}
\label{fig:Dec}
\end{figure}

\begin{figure}
\centering
\includegraphics[width=0.8\columnwidth]{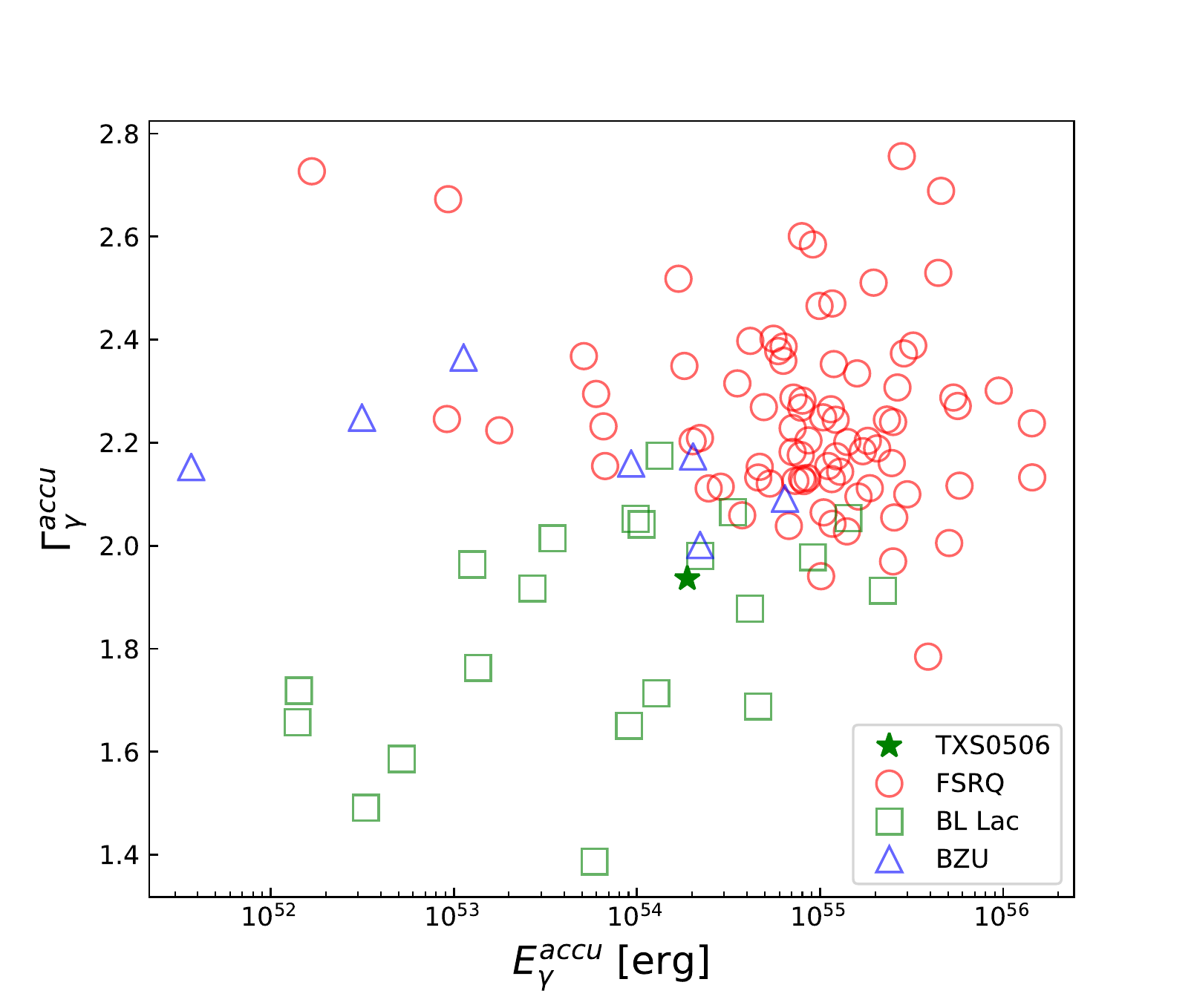}
\caption{The $E_{\gamma}^{accu}-\Gamma_\gamma^{accu}$ diagram, where $E_{\gamma}^{accu}$ and $\Gamma_\gamma^{accu}$ represent the accumulative equivalent energy and the averaged spectral index for the time bins in which the energy flux is three times larger than $f_{\gamma}^{aver}$, respectively. The legends are as the same of Fig. \ref{fig:lg}.}
\label{fig:L-hs}
\end{figure}

\begin{figure}
\centering
\includegraphics[width=0.8\columnwidth,height=0.55\textheight]{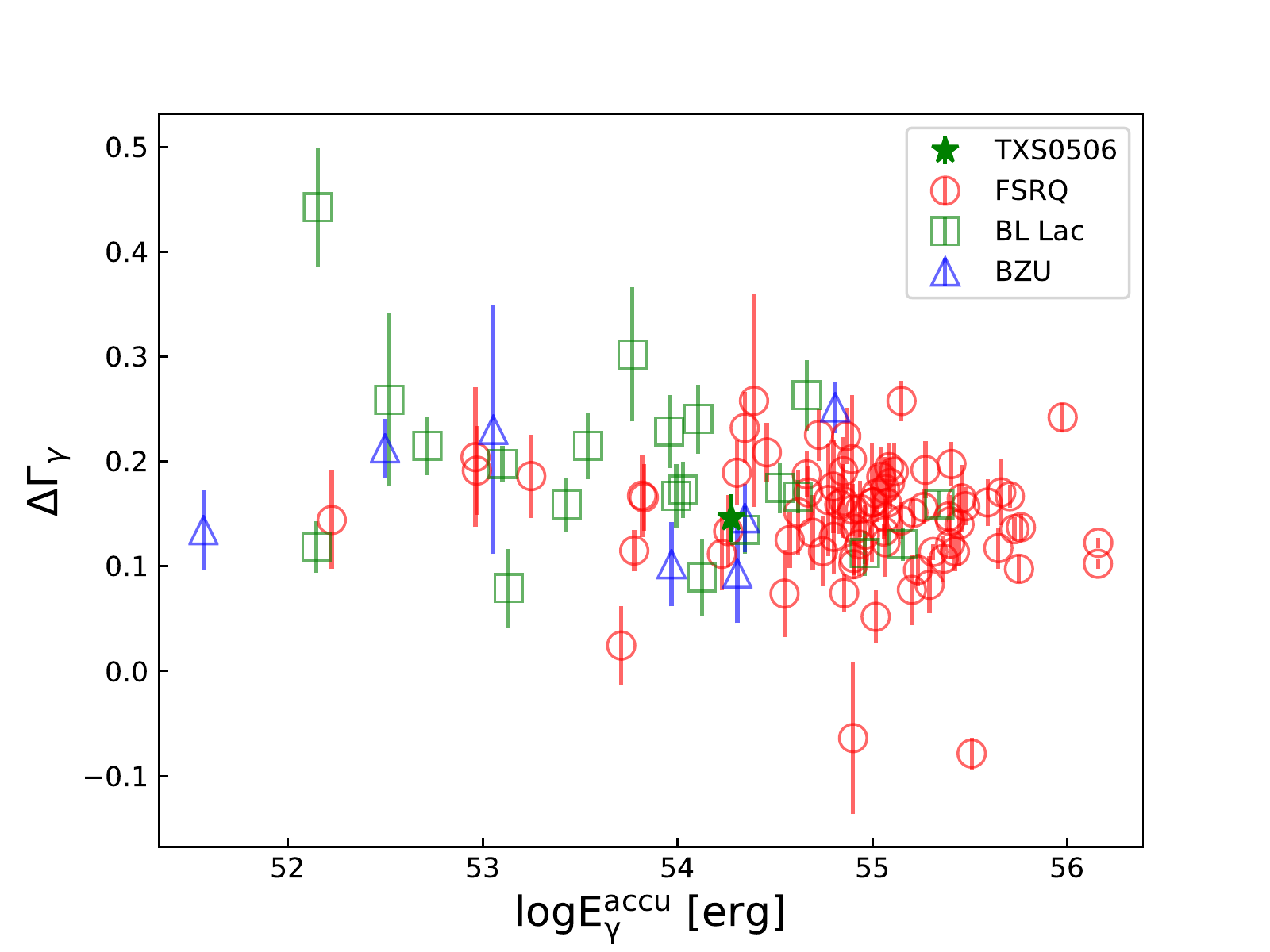}
\caption{TXS 0506+056, marked as the green pentagram, in ${\rm log} E_{\gamma}^{\bm {accu}} - \Delta \Gamma_{\gamma}$ diagram (same legend as for Fig. \ref{fig:lg}). }
\label{fig:Delta_gamma}
\end{figure}

The sources of PeV neutrinos and ultra-high-energy cosmic rays are still in debate in the literature. Thanks to the successful performance of IceCube and {\it Fermi}-LAT and the quick follow-up observations in optical, radio and X-ray, significant progresses have been made. The most exciting findings are the possible IceCube-170922A/TXS 0506+056 association \citep{IceCube2018a} and the the coincidence between a 2 PeV neutrino event and the blazar PKS B1424$-$418 \citep{2016NatPh..12..807K}, which favors the hypothesis that blazars are important sources of PeV neutrinos and ultra-high-energy cosmic rays. Motivated by the possible connection between the GeV activity and the neutrino IceCube-170922A, in this work we have analyzed the {\it Fermi}-LAT data of a group of bright AGNs and identify strong GeV flares. We have compared the properties of these bright AGNs displaying strong GeV activities with TXS 0506+056 to look for indication of the ultra-high-energy cosmic ray sources. It turns out that TXS 0506+056 appears to be similar to other bright blazars studied in this work (see Fig.\ref{fig:lg}, Fig.\ref{fig:f-l}, Fig.\ref{fig:L-hs} and Fig.\ref{fig:Delta_gamma}), except its very strong GeV activities (i.e., the high value of $f_{\gamma}^{accu}/f_{\gamma}^{aver}$ shown in Fig.\ref{fig:foc}). We suggest to carry out dedicated searches for (weak) neutrino outburst in the historical data of IceCube, as recently done for TXS 0506+056 \citep{IceCube2018b}, from the directions of these more energetic and harder sources with strong GeV activities. If null results are turned out, new indicator rather than the GeV activity should be identified for the neutrino sources. 

\acknowledgments
This research has made use of the NASA/IPAC Extragalactic Database which is operated by the
Jet Propulsion Laboratory, California Institute of Technology, under contract with the National
Aeronautics and Space Administration. This research also makes use of the SIMBAD database,
operated at CDS, Strasbourg, France.

This work was supported in part by NSFC under grants 11525313 (i.e., Funds for Distinguished Young Scholars), 11703093 and 11303098.
H.N.H. is also supported by the Special Postdoctoral Researchers (SPDR) Program in RIKEN.

\vspace{5mm}
\facilities{{\it Fermi} (LAT)}

\begin{longrotatetable}
\renewcommand{\thetable}{\arabic{table}}
\begin{deluxetable*}{cccccccccccc}
\tablenum{1}
\tablecaption{Model parameters of all target sources in the sample}\label{Table 1}
\tablewidth{0pt}
\tabletypesize{\scriptsize}
\startdata
FL8Y Name & Alias Name & R.A.$^{\,\,\,\,\text{\tt a}}$ & Dec.$^{\,\,\,\,\text{a}}$ & Photon Flux & Spectral  & Energy Flux & TS  & Redshift \\
        &            & [deg]&[deg] & [${10^{-8}}$ ph cm$^{-2}$ s$^{-1}$]&  Index &[${10^{-5}}$ MeV cm$^{-2}$ s$^{-1}$] & Value & \\
\hline 
FL8Y J0001.4+2112 & TXS 2358+209 & 0.385 & 21.227  & $ 8.747 \pm 0.237 $ & $ 2.715 \pm 0.024 $ & $ 2.090 \pm 0.048 $ & 3238 & 1.106 \\
FL8Y J0108.6+0135 & 4C +01.02 & 17.162 & 1.583  & $ 28.651 \pm 0.326 $ & $ 2.369 \pm 0.008 $ & $ 10.163 \pm 0.113 $ & 37684 & 2.099 \\
FL8Y J0112.8+3208 & 4C 31.03 & 18.210 & 32.138  & $ 4.876 \pm 0.183 $ & $ 2.392 \pm 0.025 $ & $ 1.667 \pm 0.050 $ & 2526 & 0.603 \\
FL8Y J0133.1-5201 & PKS 0131$-$522 & 23.274 & $-$52.001  & $ 3.652 \pm 0.155 $ & $ 2.287 \pm 0.026 $ & $ 1.495 \pm 0.053 $ & 2705 & 0.02 \\
FL8Y J0210.7$-$5101 & 0208$-$512 & 32.693 & $-$51.017  & $ 10.342 \pm 0.196 $ & $ 2.343 \pm 0.014 $ & $ 3.825 \pm 0.071 $ & 12026 & 1.003 \\
FL8Y J0211.2+1051 & CGRaBS J0211+1051 & 32.805 & 10.860  & $ 7.334 \pm 0.201 $ & $ 2.130 \pm 0.015 $ & $ 4.267 \pm 0.121 $ & 7080 & 0.2 \\
FL8Y J0221.1+3556 & S3 0218+35 & 35.273 & 35.937  & $ 14.846 \pm 0.283 $ & $ 2.280 \pm 0.011 $ & $ 6.157 \pm 0.100 $ & 18551 & 0.96 \\
FL8Y J0222.6+4302 & 3C 66A & 35.665 & 43.035  & $ 11.196 \pm 0.224 $ & $ 1.954 \pm 0.010 $ & $ 11.119 \pm 0.260 $ & 24508 & 0.34 \\
FL8Y J0237.9+2848 & 4C +28.07 & 39.468 & 28.802  & $ 21.721 \pm 0.198 $ & $ 2.308 \pm 0.007 $ & $ 8.554 \pm 0.097 $ & 27262 & 1.206 \\
FL8Y J0238.6+1637 & 0235+164 & 39.662 & 16.616  & $ 11.963 \pm 0.272 $ & $ 2.175 \pm 0.013 $ & $ 6.214 \pm 0.129 $ & 13023 & 0.94 \\
FL8Y J0246.0$-$4650 & PKS 0244$-$470 & 41.500 & $-$46.855  & $ 5.902 \pm 0.167 $ & $ 2.419 \pm 0.021 $ & $ 1.941 \pm 0.048 $ & 4355 & 1.385 \\
FL8Y J0252.8$-$2219 & PKS 0250$-$225 & 43.200 & $-$22.324  & $ 8.124 \pm 0.183 $ & $ 2.353 \pm 0.016 $ & $ 2.959 \pm 0.065 $ & 7095 & 1.419 \\
FL8Y J0303.4$-$2407 & PKS 0301$-$243 & 45.860 & $-$24.120  & $ 3.780 \pm 0.119 $ & $ 1.879 \pm 0.016 $ & $ 4.948 \pm 0.214 $ & 7982 & 0.26 \\
FL8Y J0324.7+3411 & 1H 0323+342 & 51.172 & 34.179  & $ 6.654 \pm 0.332 $ & $ 2.871 \pm 0.037 $ & $ 1.426 \pm 0.061 $ & 1169 & 0.0629 \\
FL8Y J0339.5$-$0146 & PKS 0336$-$01 & 54.879 & $-$1.777  & $ 15.226 \pm 0.262 $ & $ 2.346 \pm 0.012 $ & $ 5.611 \pm 0.091 $ & 13820 & 0.852 \\
FL8Y J0348.5$-$2750 & PKS 0346$-$27 & 57.159 & $-$27.820  & $ 4.332 \pm 0.158 $ & $ 2.261 \pm 0.023 $ & $ 1.865 \pm 0.061 $ & 3685 & 0.987496 \\
FL8Y J0359.6+5057 & 4C +50.11 & 59.874 & 50.964  & $ 8.821 \pm 0.342 $ & $ 2.593 \pm 0.027 $ & $ 2.351 \pm 0.076 $ & 1246 & 1.51 \\
FL8Y J0403.9$-$3605 & PKS 0402$-$362 & 60.974 & $-$36.084  & $ 18.047 \pm 0.235 $ & $ 2.539 \pm 0.011 $ & $ 5.097 \pm 0.061 $ & 21108 & 1.42284 \\
FL8Y J0428.6$-$3756 & PKS 0426$-$380 & 67.168 & $-$37.939  & $ 26.287 \pm 0.276 $ & $ 2.090 \pm 0.006 $ & $ 17.024 \pm 0.231 $ & 70329 & 1.111 \\
NONE$^{\,\,\,\,\text{\tt b}}$ & PKS 0438$-$43 & 70.072 & $-$43.552  & $ 2.221 \pm 0.237 $ & $ 2.626 \pm 0.060 $ & $ 0.573 \pm 0.044 $ & 426 & 2.852 \\
FL8Y J0457.0$-$2324 & PKS 0454$-$234 & 74.263 & $-$23.414  & $ 24.695 \pm 0.238 $ & $ 2.189 \pm 0.007 $ & $ 12.408 \pm 0.162 $ & 49810 & 1.003 \\
FL8Y J0501.2$-$0158 & PKS 0458$-$02 & 75.303 & $-$1.987  & $ 11.826 \pm 0.198 $ & $ 2.383 \pm 0.012 $ & $ 4.105 \pm 0.069 $ & 8876 & 2.285998 \\
FL8Y J0505.3+0459 & PKS 0502+049 & 76.347 & 4.995  & $ 15.895 \pm 0.390 $ & $ 2.360 \pm 0.014 $ & $ 5.720 \pm 0.103 $ & 11563 & 0.954 \\
FL8Y J0510.0+1800 & PKS 0507+17 & 77.510 & 18.012  & $ 6.096 \pm 0.306 $ & $ 2.184 \pm 0.019 $ & $ 3.002 \pm 0.118 $ & 4128 & 0.416 \\
FL8Y J0515.5$-$4556 & PKS 0514$-$459 & 78.939 & $-$45.945  & $ 3.155 \pm 0.227 $ & $ 2.410 \pm 0.038 $ & $ 1.051 \pm 0.047 $ & 1358 & 0.194 \\
FL8Y J0521.7+2112 & VER 0521+211 & 80.442 & 21.214  & $ 9.696 \pm 0.361 $ & $ 1.945 \pm 0.018 $ & $ 9.594 \pm 0.373 $ & 12723 & 2.218 \\
FL8Y J0522.9$-$3628 & PKS 0521$-$36 & 80.742 & $-$36.459  & $ 12.144 \pm 0.241 $ & $ 2.461 \pm 0.014 $ & $ 3.572 \pm 0.064 $ & 11505 & 0.056546 \\
FL8Y J0530.9+1331 & PKS 0528+134 & 82.735 & 13.532  & $ 3.921 \pm 0.321 $ & $ 2.513 \pm 0.035 $ & $ 1.183 \pm 0.064 $ & 706 & 2.06 \\
FL8Y J0532.0$-$4827 & CRATES J0531$-$4827 & 82.994 & $-$48.460  & $ 4.598 \pm 0.242 $ & $ 2.134 \pm 0.018 $ & $ 2.501 \pm 0.094 $ & 6265 & None$^{\,\,\,\,\text{\tt c}}$ \\
FL8Y J0532.7+0733 & OG 050 & 83.162 & 7.545  & $ 10.355 \pm 0.337 $ & $ 2.340 \pm 0.016 $ & $ 3.729 \pm 0.090 $ & 6004 & 1.254 \\
FL8Y J0538.8$-$4405 & PKS 0537$-$441 & 84.710 & $-$44.086  & $ 15.656 \pm 0.225 $ & $ 2.072 \pm 0.008 $ & $ 11.445 \pm 0.195 $ & 39101 & 0.894 \\
FL8Y J0622.9+3326 & B2 0619+33 & 95.718 & 33.436  & $ 5.735 \pm 0.260 $ & $ 2.244 \pm 0.020 $ & $ 2.475 \pm 0.093 $ & 3162 & None$^{\,\,\,\,\text{\tt c}}$ \\
FL8Y J0713.9+1935 & MG2 J071354+1934 & 108.482 & 19.583  & $ 2.252 \pm 0.239 $ & $ 2.595 \pm 0.057 $ & $ 0.613 \pm 0.032 $ & 348 & 0.54 \\
FL8Y J0721.9+7120 & 0716+714 & 110.430 & 71.350  & $ 22.819 \pm 0.215 $ & $ 2.044 \pm 0.005 $ & $ 14.510 \pm 0.242 $ & 94091 & 0.3 \\
FL8Y J0725.3+1425 & 4C 14.23 & 111.320 & 14.420  & $ 4.887 \pm 0.217 $ & $ 2.252 \pm 0.020 $ & $ 2.039 \pm 0.072 $ & 3421 & 1.038 \\
FL8Y J0730.3$-$1141 & PKS 0727$-$11 & 112.580 & $-$11.687  & $ 17.450 \pm 0.360 $ & $ 2.303 \pm 0.011 $ & $ 6.586 \pm 0.124 $ & 12549 & 1.591 \\
FL8Y J0739.2+0137 & PKS 0736+01 & 114.825 & 1.618  & $ 12.472 \pm 0.331 $ & $ 2.410 \pm 0.014 $ & $ 3.997 \pm 0.081 $ & 9398 & 0.18941 \\
FL8Y J0742.6+5443 & BZU J0742+5444 & 115.666 & 54.740  & $ 3.954 \pm 0.181 $ & $ 2.324 \pm 0.023 $ & $ 1.464 \pm 0.055 $ & 3314 & 0.72 \\
FL8Y J0752.0+3318 & B2 0748+33 & 117.974 & 33.222  & $ 3.561 \pm 0.167 $ & $ 2.412 \pm 0.030 $ & $ 1.183 \pm 0.043 $ & 1552 & 1.935716 \\
FL8Y J0808.2$-$0751 & PKS 0805$-$07 & 122.065 & $-$7.853  & $ 7.687 \pm 0.229 $ & $ 2.304 \pm 0.018 $ & $ 2.443 \pm 0.067 $ & 4439 & 1.837 \\
FL8Y J0830.7+2410 & 0827+243 & 127.490 & 24.220  & $ 3.856 \pm 0.270 $ & $ 2.516 \pm 0.032 $ & $ 1.160 \pm 0.053 $ & 1747 & 0.9414 \\
FL8Y J0841.5+7053 & S5 0836+71 & 130.352 & 70.895  & $ 15.342 \pm 0.253 $ & $ 2.859 \pm 0.014 $ & $ 3.286 \pm 0.046 $ & 15186 & 2.172 \\
FL8Y J0852.5$-$5755 & PMN J0852$-$5755 & 133.161 & $-$57.925  & $ 3.874 \pm 0.208 $ & $ 2.212 \pm 0.027 $ & $ 1.851 \pm 0.075 $ & 1637 & None$^{\,\,\,\,\text{\tt c}}$ \\
FL8Y J0854.8+2006 & OJ 287 & 133.704 & 20.109  & $ 8.843 \pm 0.251 $ & $ 2.214 \pm 0.014 $ & $ 3.943 \pm 0.115 $ & 10848 & 0.3056 \\
FL8Y J0904.9$-$5735 & PKS 0903$-$57 & 136.222 & $-$57.585  & $ 7.483 \pm 0.249 $ & $ 2.266 \pm 0.019 $ & $ 3.187 \pm 0.086 $ & 3829 & 0.695 \\
FL8Y J0909.1+0121 & PKS B0906+015 & 137.292 & 1.360  & $ 5.209 \pm 0.302 $ & $ 2.515 \pm 0.030 $ & $ 1.494 \pm 0.065 $ & 1909 & 1.024905 \\
FL8Y J0921.6+6216 & OK +630 & 140.401 & 62.264  & $ 5.385 \pm 0.192 $ & $ 2.334 \pm 0.019 $ & $ 1.954 \pm 0.061 $ & 5830 & 1.446 \\
FL8Y J0948.9+0022 & PMN J0948+0022 & 147.239 & 0.374  & $ 11.425 \pm 0.274 $ & $ 2.630 \pm 0.018 $ & $ 2.938 \pm 0.061 $ & 5600 & 0.58384 \\
FL8Y J0958.7+6533 & S4 0954+65 & 149.697 & 65.565  & $ 6.304 \pm 0.140 $ & $ 2.219 \pm 0.014 $ & $ 2.962 \pm 0.068 $ & 8226 & 0.367 \\
FL8Y J1006.7$-$2159 & PKS 1004$-$217 & 151.693 & $-$21.989  & $ 4.570 \pm 0.179 $ & $ 2.320 \pm 0.024 $ & $ 1.759 \pm 0.058 $ & 2556 & 0.331 \\
FL8Y J1033.9+6050 & S4 1030+61 & 158.464 & 60.852  & $ 7.498 \pm 0.234 $ & $ 2.197 \pm 0.016 $ & $ 3.699 \pm 0.090 $ & 11691 & 1.40095 \\
FL8Y J1048.4+7143 & S5 1044+71 & 162.115 & 71.727  & $ 17.398 \pm 0.240 $ & $ 2.246 \pm 0.009 $ & $ 7.715 \pm 0.105 $ & 37461 & 1.15 \\
FL8Y J1058.4+0133 & 4C +01.28 & 164.623 & 1.566  & $ 9.918 \pm 0.266 $ & $ 2.229 \pm 0.015 $ & $ 4.558 \pm 0.104 $ & 9457 & 0.18516 \\
FL8Y J1104.4+3812 & Mrk 421 & 166.114 & 38.209  & $ 21.273 \pm 0.226 $ & $ 1.776 \pm 0.005 $ & $ 42.327 \pm 0.699 $ & 117751 & 0.031 \\
FL8Y J1153.4+4930 & 1150+497 & 178.352 & 49.519  & $ 3.330 \pm 0.166 $ & $ 2.392 \pm 0.031 $ & $ 1.139 \pm 0.040 $ & 2047 & 0.33364 \\
FL8Y J1159.5+2914 & Ton 599 & 179.883 & 29.246  & $ 18.036 \pm 0.223 $ & $ 2.172 \pm 0.008 $ & $ 9.452 \pm 0.148 $ & 36503 & 0.72475 \\
FL8Y J1224.9+2122 & PKS B1222+216 & 186.227 & 21.380  & $ 36.327 \pm 0.297 $ & $ 2.390 \pm 0.007 $ & $ 12.475 \pm 0.113 $ & 65284 & 0.43383 \\
FL8Y J1229.0+0202 & 3C 273 & 187.278 & 2.052  & $ 21.798 \pm 0.364 $ & $ 2.864 \pm 0.016 $ & $ 4.692 \pm 0.065 $ & 12020 & 0.158339 \\
FL8Y J1230.2+2517 & ON 246 & 187.559 & 25.302  & $ 5.045 \pm 0.162 $ & $ 2.076 \pm 0.018 $ & $ 3.404 \pm 0.117 $ & 6680 & 0.135 \\
FL8Y J1239.5+0443 & J123939+044409 & 189.900 & 4.700  & $ 14.912 \pm 0.273 $ & $ 2.421 \pm 0.013 $ & $ 4.887 \pm 0.077 $ & 12355 & 1.761 \\
FL8Y J1246.7$-$2547 & PKS 1244$-$255 & 191.695 & $-$25.797  & $ 13.740 \pm 0.263 $ & $ 2.323 \pm 0.013 $ & $ 5.264 \pm 0.089 $ & 12886 & 0.638 \\
FL8Y J1256.1$-$0547 & 3C 279 & 194.047 & $-$5.789  & $ 80.799 \pm 0.428 $ & $ 2.303 \pm 0.004 $ & $ 32.109 \pm 0.209 $ & 196753 & 0.5362 \\
FL8Y J1312.6+4828 & GB6 B1310+4844 & 198.181 & 48.475  & $ 1.938 \pm 0.129 $ & $ 2.261 \pm 0.038 $ & $ 0.834 \pm 0.041 $ & 1205 & 0.501 \\
FL8Y J1316.1$-$3338 & PKS 1313$-$333 & 199.033 & $-$33.650  & $ 4.870 \pm 0.238 $ & $ 2.344 \pm 0.027 $ & $ 1.799 \pm 0.061 $ & 2066 & 1.21 \\
FL8Y J1332.0$-$0509 & PKS 1329$-$049 & 203.019 & $-$5.162  & $ 7.418 \pm 0.327 $ & $ 2.538 \pm 0.026 $ & $ 2.096 \pm 0.066 $ & 3105 & 2.15 \\
FL8Y J1345.5+4453 & B3 1343+451 & 206.388 & 44.883  & $ 17.004 \pm 0.194 $ & $ 2.252 \pm 0.008 $ & $ 7.449 \pm 0.107 $ & 34551 & 2.534 \\
FL8Y J1427.9$-$4206 & PKS 1424$-$41 & 216.985 & $-$42.105  & $ 64.429 \pm 0.369 $ & $ 2.172 \pm 0.008 $ & $ 33.733 \pm 0.556 $ & 168531 & 1.522 \\
FL8Y J1504.3+1029 & PKS 1502+106 & 226.104 & 10.494  & $ 24.391 \pm 0.250 $ & $ 2.254 \pm 0.008 $ & $ 10.654 \pm 0.135 $ & 40353 & 1.83928 \\
FL8Y J1506.1+3731 & B2 1504+37 & 226.540 & 37.514  & $ 6.272 \pm 0.179 $ & $ 2.482 \pm 0.021 $ & $ 1.893 \pm 0.045 $ & 4379 & 0.674 \\
FL8Y J1512.8$-$0906 & 1510$-$089 & 228.170 & $-$8.830  & $ 79.200 \pm 0.455 $ & $ 2.354 \pm 0.004 $ & $ 28.760 \pm 0.187 $ & 150184 & 0.36 \\
FL8Y J1517.7$-$2422 & AP Lib & 229.420 & $-$24.370  & $ 9.333 \pm 0.407 $ & $ 2.161 \pm 0.014 $ & $ 5.015 \pm 0.092 $ & 8952 & 0.049 \\
FL8Y J1522.1+3144 & B2 1520+31 & 230.540 & 31.740  & $ 27.550 \pm 0.162 $ & $ 2.459 \pm 0.002 $ & $ 8.570 \pm 0.037 $ & 45252 & 1.4886 \\
FL8Y J1532.7$-$1319 & TXS 1530$-$131 & 233.160 & $-$13.350  & $ 8.117 \pm 0.083 $ & $ 2.213 \pm 0.004 $ & $ 3.870 \pm 0.024 $ & 6019 & None$^{\,\,\,\,\text{\tt c}}$ \\
FL8Y J1555.7+1111 & PG 1553+113 & 238.930 & 11.190  & $ 7.017 \pm 0.072 $ & $ 1.689 \pm 0.005 $ & $ 20.460 \pm 0.316 $ & 28725 & 0.36 \\
FL8Y J1625.7$-$2527 & PKS 1622$-$253 & 246.440 & $-$25.460  & $ 10.310 \pm 0.140 $ & $ 2.301 \pm 0.003 $ & $ 4.112 \pm 0.042 $ & 3581 & 0.786 \\
FL8Y J1626.0$-$2950 & PKS B 1622$-$297 & 246.530 & $-$29.860  & $ 11.320 \pm 0.274 $ & $ 2.549 \pm 0.014 $ & $ 3.161 \pm 0.061 $ & 3426 & 0.815 \\
FL8Y J1635.2+3808 & 1633+382 & 248.810 & 38.130  & $ 31.600 \pm 0.334 $ & $ 2.360 \pm 0.007 $ & $ 11.370 \pm 0.117 $ & 56527 & 1.8131 \\
FL8Y J1640.5+3945 & 0FGL J1641.4+3939 & 250.350 & 39.670  & $ 5.175 \pm 0.094 $ & $ 2.441 \pm 0.007 $ & $ 1.650 \pm 0.021 $ & 1930 & 0.5948 \\
FL8Y J1653.8+3945 & Mrk 501 & 253.470 & 39.760  & $ 6.222 \pm 0.123 $ & $ 1.750 \pm 0.009 $ & $ 13.840 \pm 0.412 $ & 24831 & 0.033 \\
FL8Y J1700.0+6830 & GB6 J1700+6830 & 255.040 & 68.500  & $ 5.295 \pm 0.083 $ & $ 2.399 \pm 0.007 $ & $ 1.792 \pm 0.020 $ & 4206 & 0.301 \\
FL8Y J1709.7+4318 & B3 1708+433 & 257.420 & 43.310  & $ 5.185 \pm 0.067 $ & $ 2.350 \pm 0.005 $ & $ 1.898 \pm 0.016 $ & 3997 & 1.027 \\
FL8Y J1733.0$-$1304 & 1730$-$130 & 263.260 & $-$13.080  & $ 14.390 \pm 0.164 $ & $ 2.359 \pm 0.004 $ & $ 5.184 \pm 0.039 $ & 6130 & 0.902 \\
FL8Y J1734.3+3858 & OT 355 & 263.590 & 38.960  & $ 5.558 \pm 0.188 $ & $ 2.378 \pm 0.023 $ & $ 1.943 \pm 0.051 $ & 3761 & 0.975 \\
FL8Y J1748.6+7005 & S4 1749+70 & 267.140 & 70.100  & $ 3.986 \pm 0.057 $ & $ 1.951 \pm 0.006 $ & $ 3.999 \pm 0.046 $ & 8077 & 0.77 \\
FL8Y J1751.4+0938 & OT 081 & 267.890 & 9.650  & $ 8.703 \pm 0.057 $ & $ 2.264 \pm 0.002 $ & $ 3.727 \pm 0.017 $ & 5388 & 0.322 \\
FL8Y J1800.7+7828 & S5 1803+78 & 270.190 & 78.470  & $ 8.541 \pm 0.057 $ & $ 2.240 \pm 0.003 $ & $ 3.839 \pm 0.017 $ & 13318 & 0.68 \\
FL8Y J1801.4+4404 & S4 1800+44 & 270.380 & 44.070  & $ 4.391 \pm 0.078 $ & $ 2.389 \pm 0.007 $ & $ 1.509 \pm 0.019 $ & 2457 & 0.663 \\
FL8Y J1829.1$-$5814 & PKS 1824$-$582 & 277.300 & $-$58.230  & $ 10.160 \pm 0.049 $ & $ 2.593 \pm 0.002 $ & $ 2.707 \pm 0.010 $ & 4883 & 1.531 \\
FL8Y J1833.6$-$2103 & PKS 1830$-$211 & 278.420 & $-$21.060  & $ 45.950 \pm 0.001 $ & $ 2.543 \pm 0.003 $ & $ 12.910 \pm 0.001 $ & 27416 & 2.507 \\
FL8Y J1848.4+3217 & CGRaBS J1848+3219 & 282.090 & 32.320  & $ 5.727 \pm 0.019 $ & $ 2.401 \pm 0.001 $ & $ 1.933 \pm 0.005 $ & 1841 & 0.8 \\
FL8Y J1848.5+3242 & B2 1846+32B & 282.140 & 32.730  & $ 7.474 \pm 0.023 $ & $ 2.541 \pm 0.001 $ & $ 2.104 \pm 0.005 $ & 2339 & None$^{\,\,\,\,\text{\tt c}}$ \\
FL8Y J1849.2+6705 & CGRaBS J1849+6705 & 282.320 & 67.090  & $ 7.127 \pm 0.057 $ & $ 2.320 \pm 0.003 $ & $ 2.747 \pm 0.014 $ & 7111 & 0.657 \\
FL8Y J1911.2$-$2006 & PKS B1908$-$201 & 287.790 & $-$20.110  & $ 10.440 \pm 0.204 $ & $ 2.537 \pm 0.008 $ & $ 2.953 \pm 0.044 $ & 3856 & 1.119 \\
FL8Y J1958.0$-$3845 & PKS 1954$-$388 & 299.499 & $-$38.752  & $ 7.108 \pm 0.204 $ & $ 2.369 \pm 0.019 $ & $ 2.522 \pm 0.064 $ & 3783 & 0.63 \\
FL8Y J2000.0+6508 & 1ES 1959+650 & 300.000 & 65.149  & $ 6.514 \pm 0.159 $ & $ 1.801 \pm 0.011 $ & $ 11.660 \pm 0.345 $ & 19361 & 0.047 \\
FL8Y J2025.6$-$0735 & PKS 2023$-$07 & 306.419 & $-$7.598  & $ 14.490 \pm 0.234 $ & $ 2.257 \pm 0.011 $ & $ 6.293 \pm 0.109 $ & 15463 & 1.388 \\
FL8Y J2035.4+1056 & PKS 2032+107 & 308.843 & 10.935  & $ 13.090 \pm 0.318 $ & $ 2.362 \pm 0.014 $ & $ 4.693 \pm 0.091 $ & 8410 & 0.601 \\
FL8Y J2126.3$-$4606 & PKS 2123$-$463 & 321.628 & $-$46.097  & $ 3.336 \pm 0.185 $ & $ 2.535 \pm 0.038 $ & $ 0.946 \pm 0.040 $ & 953 & 1.67 \\
FL8Y J2141.6$-$6410 & PMN J2141$-$6411 & 325.003 & $-$64.026  & $ 7.776 \pm 0.197 $ & $ 2.381 \pm 0.018 $ & $ 2.708 \pm 0.063 $ & 5834 & None$^{\,\,\,\,\text{\tt c}}$ \\
FL8Y J2143.5+1743 & OX 169 & 325.898 & 17.730  & $ 14.440 \pm 0.261 $ & $ 2.450 \pm 0.014 $ & $ 4.543 \pm 0.075 $ & 9425 & 0.21074 \\
FL8Y J2147.2$-$7536 & PKS 2142$-$75 & 326.803 & $-$75.604  & $ 13.430 \pm 0.214 $ & $ 2.439 \pm 0.012 $ & $ 4.293 \pm 0.062 $ & 10384 & 1.139 \\
FL8Y J2151.9$-$3027 & PKS 2149$-$306 & 327.981 & $-$30.465  & $ 10.270 \pm 0.276 $ & $ 2.896 \pm 0.025 $ & $ 2.169 \pm 0.051 $ & 3662 & 2.345 \\
FL8Y J2158.8$-$3013 & PKS 2155$-$304 & 329.717 & $-$30.226  & $ 13.190 \pm 0.194 $ & $ 1.842 \pm 0.008 $ & $ 19.940 \pm 0.445 $ & 45844 & 0.116 \\
FL8Y J2201.8+5048 & NRAO 676 & 330.431 & 50.816  & $ 15.260 \pm 0.301 $ & $ 2.647 \pm 0.017 $ & $ 3.864 \pm 0.068 $ & 5089 & 1.899 \\
FL8Y J2202.7+4216 & BL Lac & 330.680 & 42.278  & $ 36.250 \pm 0.297 $ & $ 2.161 \pm 0.006 $ & $ 19.470 \pm 0.216 $ & 64653 & 0.0686 \\
FL8Y J2232.5+1143 & CTA 102 & 338.152 & 11.731  & $ 104.300 \pm 0.427 $ & $ 2.255 \pm 0.003 $ & $ 45.470 \pm 0.267 $ & 292708 & 1.037 \\
FL8Y J2236.5$-$1433 & PKS 2233-148 & 339.142 & $-$14.556  & $ 8.400 \pm 0.196 $ & $ 2.115 \pm 0.014 $ & $ 5.088 \pm 0.136 $ & 9629 & 0.325 \\
FL8Y J2244.2+4057 & TXS 2241+406 & 341.053 & 40.954  & $ 9.563 \pm 0.232 $ & $ 2.091 \pm 0.012 $ & $ 6.178 \pm 0.151 $ & 11734 & 1.171 \\
NONE$^{\,\,\,\,\text{\tt b}}$ & PKS 2247$-$131 & 342.498 & $-$12.855  & $ 15.480 \pm 0.451 $ & $ 2.244 \pm 0.012 $ & $ 6.902 \pm 0.175 $ & 15991 & None$^{\,\,\,\,\text{\tt c}}$ \\
FL8Y J2250.7$-$2806 & PMN J2250$-$2806 & 342.685 & $-$28.110  & $ 3.813 \pm 0.188 $ & $ 2.147 \pm 0.025 $ & $ 2.124 \pm 0.086 $ & 2908 & 0.525 \\
FL8Y J2253.9+1608 & 3C 454.3 & 343.491 & 16.148  & $ 182.800 \pm 0.554 $ & $ 2.340 \pm 0.003 $ & $ 67.980 \pm 0.273 $ & 563863 & 0.859 \\
FL8Y J2258.0$-$2759 & PKS 2255$-$282 & 344.525 & $-$27.973  & $ 9.355 \pm 0.325 $ & $ 2.422 \pm 0.019 $ & $ 3.062 \pm 0.079 $ & 6130 & 0.92584 \\
FL8Y J2311.0+3425 & B2 2308+34 & 347.772 & 34.420  & $ 8.535 \pm 0.203 $ & $ 2.352 \pm 0.016 $ & $ 3.113 \pm 0.068 $ & 6329 & 1.817 \\
FL8Y J2323.5$-$0317 & PKS 2320$-$035 & 350.883 & $-$3.285  & $ 8.531 \pm 0.255 $ & $ 2.234 \pm 0.017 $ & $ 3.880 \pm 0.098 $ & 6797 & 1.41 \\
FL8Y J2329.3$-$4956 & PKS 2326$-$502 & 352.337 & $-$49.928  & $ 23.750 \pm 0.231 $ & $ 2.232 \pm 0.007 $ & $ 10.850 \pm 0.141 $ & 48780 & 0.518 \\
FL8Y J2345.2$-$1555 & PMN J2345$-$1555 & 356.302 & $-$15.919  & $ 14.290 \pm 0.254 $ & $ 2.102 \pm 0.010 $ & $ 8.955 \pm 0.180 $ & 23111 & 0.621 \\
\hline \hline
\enddata
\end{deluxetable*}
\tablecomments{
\\
a: The coordinate (R.A., Dec.) of each source is adopted from the {\it Fermi}-LAT monitored source list\footnote{https://fermi.gsfc.nasa.gov/ssc/data/access/lat/msl\_lc/}.\\
b: Two sources have no counterparts in the FL8Y catalog which marked as ``NONE'' due to they are in a relatively quiescent state during the first eight years.\\
c: Several source marked as ``None'' have no redshift values due to the non-detection of optical spectral lines for them. 
}

\end{longrotatetable}


\end{document}